\documentstyle[aps, pra, tighten]{revtex}

\begin{document}
\draft
\hoffset= -2.5mm

\title{Coulomb screening in mesoscopic noise: a kinetic approach}

\author{Frederick Green$\dagger$
and
Mukunda P Das$\ddagger$}

\address{
$\dagger$
GaAs IC Prototyping Facility,
CSIRO Telecommunications and Industrial Physics,
PO Box 76, Epping NSW 1710, Australia
}

\address{
$\ddagger$
Department of Theoretical Physics,
Research School of Physical Sciences and Engineering,
The Australian National University,
Canberra ACT 0200, Australia
}

\maketitle

\begin{abstract}
Coulomb screening, together with degeneracy, is
characteristic of the metallic electron gas.
While there is little trace of its effects
in transport and noise in the bulk, at mesoscopic scales
the electronic fluctuations
start to show appreciable Coulomb correlations.
Within completely standard Boltzmann and Fermi-liquid
frameworks, we analyze these phenomena and their relation
to the mesoscopic fluctuation-dissipation theorem, which we prove.
We identify two distinct screening mechanisms
for mesoscopic fluctuations. One is the self-consistent
response of the contact potential in a non-uniform system.
The other couples to scattering, and is an exclusively
non-equilibrium process.
Contact-potential effects renormalize all
thermal fluctuations, at all scales.
Collisional effects are relatively short-ranged and
modify non-equilibrium noise. 
We discuss ways to detect these differences experimentally.
\end{abstract}


\section{Introduction}

In a previous paper
\cite{gdi}
(GD; see also reference \onlinecite{gdcond})
we proposed a semi-classical
kinetic theory of carrier fluctuations in metallic conductors
down to mesoscopic dimensions. Our microscopic description is based
on two canonical frameworks: the Boltzmann transport equation
and the theory of charged Fermi liquids.
Boltzmann kinetics are needed to compute non-equilibrium carrier
distributions and fluctuations, given their equilibrium
properties as input. It is Fermi-liquid theory, {\it and that
theory alone}, that can supply the necessary microscopic form
of the input.

Any inhomogeneity in a conductor
of mesoscopic size begins to show itself
even if the device is strongly metallic.
For such systems the microscopic
origins of non-equilibrium Coulomb screening have
not been examined, at least within orthodox kinetic theory.
The immediate goal of this work is to provide such a description.
Its scope should cover not just low-field phenomena,
but the technologically important high-field regime.

The formalism of GD treats the degenerate electron gas
independently of the self-consistent Coulomb fields which
accompany, and modify, fluctuations of electric charge.
This purely Fermi-liquid analysis is
related conceptually to the Lindhard approximation
\cite{pinoz};
with it, one can derive a quantitative connection
between non-equilibrium
fluctuations of the current and the rate of dissipation of
electrical energy (Joule heating).
In the bulk weak-field limit this is nothing but the
linear fluctuation-dissipation theorem;
at high fields, the connection expresses the scaling of
excess thermal noise with the ambient thermodynamic temperature.
For degenerate carriers,
such scaling is an unavoidable consequence of the asymptotic
boundary conditions for the open environment to which
the conductor is connected:
local equilibrium, and local charge neutrality.

There are other semi-classical treatments of
Coulomb screening for mesoscopic noise in metals
\cite{nagaev,naveh}.
They rely mainly on Langevin stochastics
\cite{kogancpu}
and drift-diffusion phenomenology
\cite{datta},
avoiding substantive engagement with Fermi-liquid physics;
notably, the conserving sum rules
\cite{pinoz}.
The shortcomings of diffusive noise theory are discussed elsewhere
\cite{ithaca,upon}.
By contrast, a well-defined numerical method for
Coulomb suppression has been developed by Gonz\'alez and others
\cite{reggi}
to study non-degenerate shot noise; their approach
adheres to standard Boltzmann kinetics.
We leave our own study of shot noise to a future paper.
For a preliminary account, see reference \onlinecite{gdcond}.

In the present work, our most significant result is to show
that the fluctuations of the dissipated power are
functionally distinct from the underlying microscopic
fluctuations of the current. Quite simply, each behaves
differently from the other. The consequences of this separation
will be seen only in a conductor that is
(a) metallic and (b) non-uniform. Its experimental significance, and the
corroboration of our model, are most likely to be realized in a
two-dimensional quantum channel designed to meet conditions (a) and (b).

If it is corroborated, our kinetic description will offer a
unique probe of mesoscopic phenomena that have not been
considered to date. Moreover, these effects can be analyzed
equally well in the regime of {\it non-linear}
transport and noise. That is the import of our work.
The fact that it treats, simultaneously, the disparate
issues of non-uniform Coulomb screening, degeneracy and
non-linearity requires an extended technical commitment.

A prime example of a system with strong inbuilt screening effects
is the two-dimensional electron gas, self-consistently confined
in the quantum well of a III-V heterojunction
\cite{vinter};
we examine how to set up a first-principles kinetic
description for these significant systems, as well as others.
In section 2 we revisit the kinetic formalism of GD
and the structure of its Green functions.
The analysis is then extended, in section 3, to fluctuations
in the presence of self-consistent Coulomb screening;
the task divides neatly into a part governed by the
equilibrium properties of an inhomogeneous
conductor, and a part governed purely by non-equilibrium scattering.
Section 4 applies the screening formalism to the proof of
the fluctuation-dissipation (or Johnson-Nyquist) relation 
for inhomogeneous mesoscopic systems.
We discuss the difference between dissipation and fluctuations
in the non-linear regime. This leads
into section 5, in which we describe a new class of
experiments by which these differences could be measured. 
In section 6 we summarize.

\section{Kinetic formalism}

\subsection{Single-particle transport}

We review the structure of our kinetic model,
before systematizing its self-consistent
Coulomb corrections. The equation of motion
for the fluctuations is derived by variational analysis of
the semi-classical Boltzmann transport
equation for the electron distribution function
$f_{\alpha} (t) \equiv f_{s}({\bf r}, {\bf k}, t)$.
This is

\begin{equation}
{\left( {\partial\over {\partial t}}
+ D_{\alpha}[{\bf E}({\bf r},t)] \right)}
f_{\alpha}(t)
= -{\cal W}_{\alpha}[f].
\label{AX1}
\end{equation}

\noindent
The notation is as follows. Labels $\alpha = \{{\bf r}, {\bf k}, s\}$,
$\alpha' = \{{\bf r'}, {\bf k'}, s'\}$ and so on will
denote points in single-particle phase space. A sub-label {\it s}
will index both the discrete sub-bands (or valleys) of
a multi-level system and the spin state.
The system is acted upon by the total
internal field ${\bf E}({\bf r},t)$, entering via
the convective operator

\[
D_{\alpha}[{\bf E}] \equiv
{\bf v}_{\bf k}{\bbox \cdot} { {\partial}\over {\partial {\bf r} }}
- { { e{\bf E}}\over \hbar}{\bbox \cdot}
  { {\partial}\over {\partial {\bf k}} }.
\]

\noindent
The collision operator ${\cal W}_{\alpha}[f]$ may be for
any combination of single-particle impurity scattering
and two-particle scattering. Its kernel is local in real space
and satisfies detailed balance. (Pauli blocking of
the outgoing scattering states means that ${\cal W}$ is
at least a quadratic functional of $f$.)
In a system with $\nu$ dimensions, the identity operator is

\[
{\cal I}_{\alpha \alpha'}
\equiv \delta_{s  s'}
{\left\{ { {\delta_{{\bf r} {\bf r'}}}\over
	   {\Omega({\bf r})} } \right\}}
{\left\{ \Omega({\bf r}) \delta_{{\bf k} {\bf k'}} \right\}}
\longleftrightarrow \delta_{s  s'}
\delta({\bf r} - {\bf r'})
(2\pi)^{\nu} \delta({\bf k} - {\bf k'}).
\]

At equilibrium, equation (\ref{AX1}) is

\begin{equation}
D_{\alpha}[{\bf E}_0({\bf r})] f^{\rm eq}_{\alpha}(t)
= 0 = -{\cal W}_{\alpha}[f^{\rm eq}].
\label{AX1.1}
\end{equation}

\noindent
The internal field ${\bf E}_0({\bf r})$, defined in the absence of a
driving field, satisfies

\begin{equation}
{\partial\over {\partial {\bf r}} } {\bbox \cdot}
\epsilon {\bf E}_0 = -4\pi e
{\Bigl( 
{\langle f^{\rm eq}({\bf r}) \rangle} - n_{\rm d}({\bf r})
\Bigr)}
\label{poissoneq}
\end{equation}

\noindent
in terms of the dielectric constant $\epsilon({\bf r})$,
the electron density
$\langle f^{\rm eq}({\bf r}) \rangle \equiv
{\Omega({\bf r})}^{-1}{\sum}_{{\bf k},s}
f^{\rm eq}_{\alpha}$,
and the positive background density $n_{\rm d}({\bf r})$.
The mean total particle number within the conductor is
$\sum_{\bf r} \Omega({\bf r}) {\langle f^{\rm eq}({\bf r}) \rangle} = N$
and the equilibrium function is

\begin{equation}
f^{\rm eq}_{\alpha} ~=~
{\left[
1 + \exp \!
 {\left(
   { {\varepsilon_{\alpha} - \phi_{\alpha}}\over k_{B}T }
  \right)}
\right]}^{-1}.
\label{AX1.2}
\end{equation}

\noindent
Here the local conduction-band energy
$\varepsilon_{\alpha} = \varepsilon_s({\bf k}; {\bf r})$
can have band parameters that depend {\it implicitly} on position
($\partial /\partial {\bf r}$ does not act upon
$\varepsilon_{\alpha}$).
The local Fermi level
$\phi_{\alpha} = \mu - V_0({\bf r})$ is the difference
of the global chemical potential $\mu$ and the
electrostatic potential $V_0({\bf r})$,
whose gradient is $e{\bf E}_0({\bf r})$.
Both $V_0$ and ${\bf E}_0$ must vanish asymptotically
in the macroscopic leads carrying current into and out
of the mesoscopic device.

In steady state, the non-equilibrium problem can be solved for
the difference $g_{\alpha} = f_{\alpha} - f^{\rm eq}_{\alpha}$.
After subtracting both sides of equation (\ref{AX1.1}) from those
of equation (\ref{AX1}), the equation for $g$ is

\begin{equation}
\sum_{\alpha'}
{\Bigl( {\cal I}_{\alpha \alpha'}
D_{\alpha'}[{\bf E}({\bf r'})]
+ {\cal W}^{(1)}_{\alpha \alpha'}[f]
\Bigr)} g_{\alpha'}
=
{ { e{\widetilde {\bf E}}({\bf r}) }\over \hbar}{\bbox \cdot}
{ {\partial f^{\rm eq}_{\alpha}}\over {\partial {\bf k}} }
- {\cal W}^{(2)}_{\alpha}[g].
\label{AX3}
\end{equation}

\noindent
This contains the difference field
${\widetilde {\bf E}}({\bf r}) = {\bf E}({\bf r}) - {\bf E}_0({\bf r})$
and two auxiliary collision operators.
The operator ${\cal W}_{\alpha \alpha'}^{(1)}[f] \equiv
\delta {\cal W}_{\alpha}[f]/\delta f_{\alpha'}$ is the linearized
term. On the right-hand side we have the residual non-linear part

\[
{\cal W}^{(2)}_{\alpha}[g]
= {\cal W}_{\alpha}[f] - {\cal W}_{\alpha}[f^{\rm eq}]
- \sum_{\alpha'}{\cal W}^{(1)}_{\alpha \alpha'}[f]g_{\alpha'}.
\]

\noindent
For elastic impurity scattering, ${\cal W}^{(2)} = 0$; for inelastic
one-body processes, ${\cal W}^{(2)}$ is bilinear in $g$.
Note that, although ${\cal W}[f^{\rm eq}]$ is zero, its functional
derivative ${\cal W}^{(1)}[f^{\rm eq}]$ does not vanish identically.
Hence we must carry ${\cal W}[f^{\rm eq}]$ formally in equation
(\ref{AX3}), since its variation is needed below.

Next, write the difference field as
${\widetilde {\bf E}}({\bf r})
= {\bf E}_{\rm ext}({\bf r}) + {\bf E}_{\rm ind}({\bf r})$.
The first term ${\bf E}_{\rm ext}$
is the external driving field, while
${\bf E}_{\rm ind}$ is the induced field
whose Poisson equation is

\begin{equation}
{\partial\over {\partial {\bf r}} } {\bbox \cdot}
\epsilon {\bf E}_{\rm ind}
= -4\pi e
{\Bigl( \langle f({\bf r}) \rangle
- \langle f^{\rm eq}({\bf r}) \rangle \Bigr)}
= -4\pi e \langle g ({\bf r}) \rangle.
\label{poiss}
\end{equation}

\noindent
This assumes that the background charge density
$n_{\rm d}({\bf r})$ is independent of the driving field.
The non-equilibrium solution
$g$ has two properties. From equation (\ref{AX3}) one sees that
it is generated by an inhomogeneous driving term,
dependent on the equilibrium state through the factor
$\partial f^{\rm eq}_{\bf k}/\partial {\bf k}$.
Moreover, equation (\ref{poiss}) ensures that $g$
goes to zero with ${\widetilde {\bf E}}$, so that the
adiabatic connection of the steady state
to the equilibrium state is maintained.

As with equation (\ref{AX1.1}), the solution to equation
(\ref{AX3}) complies with two asymptotic conditions
for the source and drain reservoirs:
(i) {\it local equilibrium} and
(ii) {\it local charge balance}
(in three dimensions this means strict neutrality,
while in a two-dimensional quantum-confined structure this means
that the remote ionized donor layer stabilizes the
confined electron gas).
The active region must include the
carriers in the boundary layers between the conductor
and its source and drain, out to several
screening lengths. Beyond this, the internal fields are
negligible. For numerical convenience one shorts out the fields
so that, within the operative channel geometry,
${\bf E}({\bf r}) = {\bf E}_0({\bf r}) = {\bf 0}$
outside the boundaries of the device.
Gauss's theorem implies that the active
device remains globally neutral:

\begin{equation}
\sum_{\bf r} \Omega({\bf r})\langle g ({\bf r}) \rangle
\equiv \sum_{\alpha} g_{\alpha} = 0.
\label{gauss}
\end{equation}

\noindent
Hence $\sum_{\alpha} f_{\alpha} = \sum_{\alpha} f^{\rm eq}_{\alpha} = N$.
The mean total carrier number is constant.

\subsection{Adiabatic fluctuations}

Non-equilibrium electron-hole fluctuations in the steady state
are built up through the adiabatic propagator
\cite{fg1}

\begin{equation}
G_{\alpha \alpha'} \buildrel \rm def \over =
{ { \delta g_{\alpha} }\over
  { \delta f^{\rm eq}_{\alpha'} } }
\label{AX6.1}
\end{equation}

\noindent
with a global constraint coming directly
from equation (\ref{gauss}):

\begin{equation}
\sum_{\alpha} G_{\alpha \alpha'} = 0.
\label{dgauss}
\end{equation}

\noindent
The equation for $G$ follows by taking variations with respect
to $f^{\rm eq}$ on both sides of equation (\ref{AX3}). We obtain

\begin{equation}
\sum_{\beta}
{\Bigl( {\cal I}_{\alpha \beta}
D_{\beta}[{\bf E}({\bf r}_{\beta})]
+ {\cal W}^{(1)}_{\alpha \beta}[f]
\Bigr)} G_{\beta \alpha'}
= {\cal I}_{\alpha \alpha'}
{ { e{\bf {\widetilde E}}({\bf r'})}\over {\hbar}}{\bbox \cdot}
   { {\partial }\over {\partial {\bf k'}} }
- {\cal W}^{(1)}_{\alpha \alpha'}[f]
+ {\cal W}^{(1)}_{\alpha \alpha'}[f^{\rm eq}].
\label{AXG}
\end{equation}

\noindent
The variation is restricted to exclude the reaction of
${\bf E}_0$ and ${\bf E}$. Thus $G$ is a ``proper''
response function, free of Coulomb screening.
Here we treat the electrons as an effectively neutral Fermi liquid.
Section 3 below has the complete fluctuation structure,
including Coulomb effects.

One of the most important features of
equation (\ref{AXG}) is its validity for both single-particle
and two-particle scattering.
This can be checked by direct expansion of its right-hand
collision terms, given an elastic two-body kernel.
It means that the whole kinetic noise formalism is
immediately applicable to semi-classical electron-electron scattering.

{\it All} of the physical consequences of the theory will therefore
hold equally well when electron-electron collisions are significant.
The most notable such consequence is the overall thermal-temperature
scaling of non-equilibrium fluctuations for degenerate electrons.

The operator $G$ acts upon the equilibrium Fermi-liquid
fluctuations. These are electron-hole pair excitations;
in our model they are given by the static long-wavelength limit
of the free-electron polarization function
\cite{pinoz},
normalized by the thermal energy:

\begin{equation}
\Delta f^{\rm eq}_{\alpha} \equiv
k_{\rm B} T { {\partial f^{\rm eq}_{\alpha} }
      \over {\partial \phi_{\alpha}} }
= f^{\rm eq}_{\alpha}(1 - f^{\rm eq}_{\alpha}).
\label{AX7.0}
\end{equation}

\noindent
Now consider the two-point electron-hole correlation
$\Delta f^{(2)}_{\alpha \alpha'} \equiv
( {\cal I}_{\alpha \alpha'} + G_{\alpha \alpha'} )
\Delta f^{\rm eq}_{\alpha'}$. The trace over $\alpha'$
of this elementary, non-equilibrium pair excitation is

\begin{equation}
\Delta f_{\alpha}
= \sum_{\alpha'} \Delta f^{(2)}_{\alpha \alpha'}
= \Delta f^{\rm eq}_{\alpha}
+ \sum_{\alpha'} G_{\alpha \alpha'} \Delta f^{\rm eq}_{\alpha'},
\label{AX7}
\end{equation}

\noindent
which is readily shown to be an exact solution to the
linearized Boltzmann equation:

\begin{equation}
\sum_{\beta} {\Bigl(
{\cal I}_{\alpha \beta}D_{\beta}[{\bf E}({\bf r}_{\beta})]
+ {\cal W}^{(1)}_{\alpha \beta}[f] \Bigr)} {\Delta f}_{\beta} = 0.
\label{AX7.B}
\end{equation}

\noindent
Once the explicit form of $G$ is known,
the thermal fluctuation structure
of the steady state is completely specified
by equation (\ref{AX7}). The most important implication of
equations (\ref{dgauss}) and (\ref{AX7}) is
conservation of the total fluctuation strength
over the conductor. Thus

\begin{equation}
\sum_{\alpha} \Delta f_{\alpha}
= \sum_{\alpha} \Delta f^{\rm eq}_{\alpha}
\equiv \Delta N.
\label{DN}
\end{equation}

\noindent
This is the direct consequence of asymptotic neutrality and equilibrium.
Our next task is to show how the dynamical Boltzmann equation indeed
engenders $G$, and hence the explicit form of the adiabatic fluctuations.

\subsection{Dynamic fluctuations}

The adiabatic distribution $\Delta f$ is the average strength
of non-equilibrium electron-hole excitations generated by spontaneous
energy exchange with the thermal bath, say at time $t'$.
The evolution of these spontaneous pair processes
is governed by the inhomogeneous time-dependent Boltzmann equation

\begin{equation}
\sum_{\beta} {\left\{
{\cal I}_{\alpha \beta}
{\left( {\partial\over {\partial t}} +
D_{\beta}[{\bf E}({\bf r}_{\beta})]
\right)}
+ {\cal W}^{(1)}_{\alpha \beta}[f]
\right\}}
R_{\beta \alpha'}(t - t')
= \delta(t - t') {\cal I}_{\alpha \alpha'}
\label{AX8.0}
\end{equation}

\noindent
for the retarded Green function

\begin{equation}
R_{\alpha \alpha'} (t - t')
\buildrel \rm def \over =
\theta(t - t')
{{\delta f_{\alpha}(t)}\over {\delta f_{\alpha'}(t')}}
\label{AXR}
\end{equation}

\noindent
with initial value
$R_{\alpha \alpha'}(0) = {\cal I}_{\alpha \alpha'}$.
As with $G$, the variation is restricted.
The double-time fluctuation

\begin{equation}
\Delta f^{(2)}_{\alpha \alpha'}(t - t')
= R_{\alpha \alpha'}(t - t') \Delta f_{\alpha'}
\label{AXR.0}
\end{equation}

\noindent
carries information on dynamic electron-hole processes
in the driven system. It is the basis for all the
correlations of physical interest. Formally,
$\Delta f^{(2)}_{\alpha \alpha'}(t - t')$
represents the strength of a fluctuation at point ${\alpha}$
at time $t$ following an initial, spontaneous excitation
at $({\alpha'}, t')$ whose strength is
$\Delta f_{\alpha'}$.

In the frequency domain, the Fourier-transformed propagator
${\cal R}(\omega) = \int dt e^{i\omega t} R(t)$
has the equation of motion

\begin{equation}
\sum_{\beta} {\cal B}_{\alpha \beta}(\omega)
{\cal R}_{\beta \alpha'}(\omega)
\equiv \sum_{\beta} {\left\{
{\cal I}_{\alpha \beta}
( D_{\beta}[{\bf E}({\bf r}_{\beta})] - i\omega )
+ {\cal W}^{(1)}_{\alpha \beta}[f]
\right\}}
{\cal R}_{\beta \alpha'}(\omega) = {\cal I}_{\alpha \alpha'},
\label{AXX}
\end{equation}

\noindent
where ${\cal B}(\omega)$ is the linearized dynamical
Boltzmann operator (shown in full in the middle line),
for which ${\cal R}(\omega)$
is the inverse.
The normalization of ${\cal R}(\omega)$ is conserved:
$\sum_{\alpha} {\cal R}_{\alpha \alpha'}(\omega)
= -1/i(\omega + i0^+)$.
This can be established from equation (\ref{AXX})
taken together with its adjoint
\cite{kogan}.
Comparison of equation (\ref{AXX}) with
equation (\ref{AX8.0}) in its adiabatic $t \to \infty$
limit shows that the low-frequency form of
$R_{\alpha \alpha'}(\omega)$ must be

\begin{equation}
{\cal R}_{\alpha \alpha'}(\omega)
\to -{1\over {i(\omega + i0^+)} }
{{\Delta f_{\alpha}}\over {\Delta N}}.
\label{rgf5}
\end{equation}

\noindent
This asymptote retains no memory of the initial state $\alpha'$. 

To make the connection between the adiabatic response $G$
in the system and the dynamic one, ${\cal R}$,
we start from an identity for
the right-hand side of equation ({\ref{AXG}):
 
\begin{eqnarray}
{\cal I}_{\alpha \alpha'}
{ { e{\bf {\widetilde E}}({\bf r'})}\over \hbar}{\bbox \cdot}
   { {\partial }\over {\partial {\bf k'}} }
- {\cal W}^{(1)}_{\alpha \alpha'}[f]
+ {\cal W}^{(1)}_{\alpha \alpha'}[f^{\rm eq}].
=&&
{\{ {\cal I}_{\alpha \alpha'}
(D_{\alpha'}[{\bf E}_0({\bf r'})] - i\omega)
+ {\cal W}^{(1)}_{\alpha \alpha'}[f^{\rm eq}] \}}
\cr
&&
- {\{ {\cal I}_{\alpha \alpha'}
(D_{\alpha'}[{\bf E}({\bf r'})] - i\omega)
+ {\cal W}^{(1)}_{\alpha \alpha'}[f] \}}
\cr
{\left. \right.} \cr
=&&
{\cal B}^{\rm eq}_{\alpha \alpha'}(\omega)
- {\cal B}_{\alpha \alpha'}(\omega).
\label{opid}
\end{eqnarray}

\noindent
The object ${\cal B}^{\rm eq}(\omega)$ is the
linearized equilibrium operator.
We can continue equation ({\ref{AXG}) into the frequency domain,
introducing the operator ${\cal G}(\omega)$ as the solution to

\begin{equation}
{\cal B}(\omega) \cdot {\cal G}(\omega)
= {\cal B}^{\rm eq}(\omega) - {\cal B}(\omega),
\label{AXG.1}
\end{equation}

\noindent
where we adopt an abbreviated notation for inner products.
One way of expressing the solution is

\begin{equation}
{\cal R}(\omega) = [{\cal I} + {\cal G}(\omega)]
\cdot {\cal R}^{\rm eq}(\omega).
\label{AXR.2}
\end{equation}

\noindent
In the zero-frequency limit, one can equate residues
at the pole $\omega = 0$ to get

\begin{equation}
{{\Delta f}\over {\Delta N}}
= [{\cal I} + {\cal G}(0)]
\cdot {{\Delta f^{\rm eq}}\over {\Delta N}}.
\label{AXR.3}
\end{equation}

\noindent
This is equation (\ref{AX7}) with $G$ identified as ${\cal G}(0)$.

To determine the form of ${\cal G}(\omega)$,
we solve equation (\ref{AXG.1}) differently:

\begin{equation}
{\cal G}(\omega) = {\cal R}(\omega) \cdot
{\left( {\cal I}
{ { e{\bf {\widetilde E}}({\bf r'})}\over {\hbar}}{\bbox \cdot}
   { {\partial }\over {\partial {\bf k'}} }
- {\cal W}^{(1)}[f] + {\cal W}^{(1)}[f^{\rm eq}]
\right)}.
\label{AXG.2}
\end{equation}

\noindent
The resolvent ${\cal R}(\omega)$ can be split into its
proper adiabatic part, dominant at low frequency, and a correlated part
${\cal C}_{\alpha \alpha'}(\omega)$ which expresses
all of the dynamics:

\begin{equation}
{\cal R}_{\alpha \alpha'}(\omega)
= {\cal C}_{\alpha \alpha'}(\omega)
-{1\over {i(\omega + i0^+)}}
{{\Delta f_{\alpha}}\over {\Delta N}}.
\label{Ccorr}
\end{equation}

\noindent
In the frequency domain, the correlated propagator ${\cal C}$
satisfies the pair of sum rules
\cite{kogan}

\begin{mathletters}
\label{cckt}
\begin{equation}
\sum_{\alpha'} {\cal C}_{\alpha \alpha'}(\omega)
\Delta f_{\alpha'} = 0
{~~~~} {\rm for {~} all} {~} \alpha,
\label{cckta}
\end{equation}

\begin{equation}
\sum_{\alpha} {\cal C}_{\alpha \alpha'}(\omega) = 0
{~~~~} {\rm for {~} all} {~} \alpha'.
\label{ccktb}
\end{equation}
\end{mathletters}

\noindent
With these we can resume the calculation of ${\cal G}(\omega)$.
The adiabatic part of ${\cal R}(\omega)$ makes no contribution to
the right-hand sum in equation (\ref{AXG.2});
it decouples both from the trace
${\langle \partial F/\partial {\bf k'} \rangle}'$, which vanishes
over the space of physical distributions $F_{\alpha'}$,
and from $\sum_{\beta} {\cal W}^{(1)}_{\beta \alpha'}$,
which is identically zero. Since
${\cal C}(\omega \to 0)$ is regular and well-defined, we finally have

\begin{equation}
G_{\alpha \alpha'}
= {\cal G}_{\alpha \alpha'}(0)
= \sum_{\beta} {\cal C}_{\alpha \beta}(0)
{\left(
{\cal I}_{\beta \alpha'}
{ { e{\bf {\widetilde E}}({\bf r'})}\over {\hbar}}{\bbox \cdot}
   { {\partial }\over {\partial {\bf k'}} }
- {\cal W}_{\beta \alpha'}^{(1)}[f] + {\cal W}_{\beta \alpha'}^{(1)}[f^{\rm eq}]
\right)}.
\label{AXG.3}
\end{equation}

\noindent
This result completes the steady-state description.
Alongside equation (\ref{AXR.2}) it shows the
intimate connection between adiabatic fluctuations,
via $G = {\cal G}(0)$, and transient ones, via ${\cal C}$.
The steady-state fluctuation structure is one with the dynamics.

\section{Coulomb correlations}

So far we have revisited three cardinal ideas:
\smallskip

(a) the governing role of asymptotic neutrality and equilibrium;
\smallskip

(b) the form of the non-equilibrium electron-hole
fluctuations as linear functionals of the equilibrium ones; and
\smallskip

(c) the analytic connection between dynamics and adiabaticity.
\smallskip

\noindent
We now apply these principles to mesoscopic conductors
as Coulomb systems, to obtain the following properties:
\smallskip

(1) renormalization of equilibrium fluctuations by long-range screening;
\smallskip

(2) short-ranged collisional screening away from equilibrium; and
\smallskip

(3) the full form of the current-current correlation function.
\smallskip

\subsection{Equilibrium screening}

We start by calculating Coulomb screening
within the conductor at equilibrium.
In a {\it correlated} system, the total fluctuation is

\begin{equation}
{\widetilde \Delta} f^{\rm eq}_{\alpha}
= k_{\rm B}T {{\delta f^{\rm eq}_{\alpha}}\over {\delta \mu}}
= {{\delta \phi_{\alpha}}\over {\delta \mu}}
\Delta f^{\rm eq}_{\alpha},
\label{TF1}
\end{equation}

\noindent
where we have used equation (\ref{AX1.2})
for $f^{\rm eq}_{\alpha}$.
Freezing the response of the internal field
means that $\delta \phi_{\alpha}/\delta \mu = 1$.
Inclusion of the self-consistent response means that

\begin{equation}
{{\delta \phi_{\alpha}}\over {\delta \mu}}
= 
1 - {{\delta V_0({\bf r})}\over {\delta \mu}}
= 1 - \sum_{\bf r'} \Omega({\bf r'})
V_{\rm C}({\bf r} - {\bf r'})
{\delta\over {\delta \mu}}
{\Bigl(
{\langle f^{\rm eq}({\bf r'}) \rangle} - n_{\rm d}({\bf r'})
\Bigr)},
\label{TF2}
\end{equation}

\noindent
where $V_{\rm C}({\bf r}) = e^2/\epsilon |{\bf r}|$ is the
Coulomb interaction. Equation (\ref{TF2}) follows from the
integral form of equation (\ref{poissoneq}).

We must address the response of the neutralizing background
$n_{\rm d}({\bf r})$. Asymptotically, the source and drain
regions are {\it unconditionally} neutral. This requires
that $n_{\rm d}({\bf r})
\equiv {\langle f^{\rm eq}({\bf r}) \rangle}$
always and everywhere outside the conductor
\cite{openbc}.
Beyond the sample boundaries, $V_{\rm C}$
is completely screened out. Thus the contribution of the
integrand on the right-hand side of
equation (\ref{TF2}) extends over the active region $\Omega$.

Within the conductor as such, the behaviour of $n_{\rm d}$ involves
physical processes outside the kinetic description.
If the background distribution is taken as independent
of transport, it is sensitive only to the
equilibrium carrier distribution within $\Omega$.
The corresponding contribution to equation (\ref{TF2}) is

\[
{ {\delta n_{\rm d}({\bf r'})}\over {\delta f^{\rm eq}_{\alpha''}} }
{ {{\widetilde \Delta} f^{\rm eq}_{\alpha''}}\over k_{\rm B}T}.
\]

\noindent
This lets us recast equation (\ref{TF2}) as

\begin{equation}
{{\delta \phi_{\alpha}}\over {\delta \mu}}
\equiv 1 - \sum_{\alpha''} {\widetilde V}_{\rm C}({\bf r}, {\bf r''})
{{{\widetilde \Delta} f^{\rm eq}_{\alpha''}}\over k_{\rm B}T}
\label{TF3}
\end{equation}

\noindent
in terms of the effective interaction

\begin{equation}
{\widetilde V}_{\rm C}({\bf r}, {\bf r''})
= V_{\rm C}({\bf r} - {\bf r''})
-\sum_{\bf r'} \Omega({\bf r'}) V_{\rm C}({\bf r} - {\bf r'})
{{\delta n_{\rm d}({\bf r'})} \over {\delta f^{\rm eq}_{\alpha''}}}.
\label{TF4}
\end{equation}

\noindent
In the most common case, the background $n_{\rm d}$ does not
change (the rigid jellium model), so that there is
no compensatory feedback from background screening:
${\widetilde V}_{\rm C} = V_{\rm C}$.
In the opposite extreme we have
${\delta n_{\rm d}}({\bf r'})/\delta f^{\rm eq}_{\alpha''}
= \delta_{{\bf r'}{\bf r''}}/\Omega({\bf r'})$,
leading to complete cancellation of the local
field generated by the carrier fluctuations.
We will assume {\it at most} an overall dependence on $N$.
That is, $\delta n_{\rm d}/\delta f^{\rm eq}_{\alpha'}
= \delta n_{\rm d}/\delta N$ for all $\alpha'$.
This applies to cases such
as the modulation-doped heterojunction
\cite{vinter},
an important example of a structure whose
background donors and active carriers
are coupled via $N$, but with incomplete compensation
owing to the physical separation of carriers and donors.

A closed form for ${\widetilde \Delta} f^{\rm eq}$
can be obtained by combining equations (\ref{TF1}) and (\ref{TF3}).
Rather than follow that course, it is more revealing
to analyze the problem kinetically.

\subsection{Boltzmann formulation of screening}

The kinetic equation for ${\widetilde \Delta} f^{\rm eq}$
is obtained by operating on equation (\ref{AX1.1})
for the equilibrium distribution. We have

\begin{equation}
\sum_{\alpha'} {\cal B}^{\rm eq}_{\alpha \alpha'}(0)
{\widetilde \Delta} f^{\rm eq}_{\alpha'}
-
{\left(
{e \over \hbar} \sum_{\alpha'}
{ {\delta {\bf E}_0({\bf r})}\over {\delta f^{\rm eq}_{\alpha'}} }
{\widetilde {\Delta}} f^{\rm eq}_{\alpha'}
\right)}
{\bbox \cdot} 
{{\partial f^{\rm eq}_{\alpha} }\over {\partial {\bf k}} }
= 0.
\label{scr5}
\end{equation}

\noindent
The second term on the left-hand side is the response of the
self-consistent field.
This equation is a reformulation of equation (\ref{TF1});
the distribution ${\widetilde \Delta}f^{\rm eq}$ given
by the latter satisfies equation (\ref{scr5})
when detailed balance is used to eliminate the
collisional contribution, giving
${\cal B}^{\rm eq}(0) \cdot {\widetilde \Delta} f^{\rm eq}
= D[{\bf E}_0] {\widetilde \Delta} f^{\rm eq}$.
However, the specifically kinetic structure of equation (\ref{scr5})
provides a rather different window on the screening physics.

Equation (\ref{poissoneq}) implies that,
within $\Omega$, the variation of $e{\bf E}_0$ with respect to$f^{\rm eq}$
is the effective electrostatic force due to an electron:

\begin{equation}
e{ {\delta {\bf E}_0({\bf r})}
\over {\delta f^{\rm eq}_{\alpha'}} }
\equiv -e{\widetilde {\bf E}}_{\rm C}({\bf r}, {\bf r'})
= {\partial\over {\partial {\bf r}}}
{\widetilde V}_{\rm C}({\bf r}, {\bf r'}).
\label{scr0}
\end{equation}

\noindent
Consequently equation (\ref{scr5}) becomes

\begin{equation}
\sum_{\alpha'} {\cal B}^{\rm eq}_{\alpha \alpha'}(0)
{\widetilde \Delta} f^{\rm eq}_{\alpha'}
= -{e \over \hbar}
{{\partial f^{\rm eq}_{\alpha} }\over {\partial {\bf k}} }
{\bbox \cdot}  \sum_{\alpha'}
{\widetilde {\bf E}}_{\rm C}({\bf r}, {\bf r'})
{\widetilde {\Delta}} f^{\rm eq}_{\alpha'}.
\eqnum{$\ref{scr5}'$}
\end{equation}

\noindent
As a variant of the equilibrium Boltzmann equation,
equation ({$\ref{scr5}'$}) is inhomogeneous. Its general
solution includes a term
proportional to the homogeneous solution,
the proper fluctuation $\Delta f^{\rm eq}$.
Let $\gamma_{\rm C}$ be the proportionality constant. Then

\begin{equation}
{\widetilde {\Delta}} f^{\rm eq}_{\alpha}
= \gamma_{\rm C} \Delta f^{\rm eq}_{\alpha}
- {e\over \hbar}\sum_{\beta}
{\cal C}^{\rm eq}_{\alpha \beta}(0)
{ {\partial f^{\rm eq}_{\beta}}
\over {\partial {\bf k}_{\beta}} } {\bbox \cdot}
\sum_{\alpha'}
{\widetilde {\bf E}}_{\rm C}({\bf r}_{\beta}, {\bf r'})
{\widetilde {\Delta}} f^{\rm eq}_{\alpha'},
\label{scr1.1}
\end{equation}

\noindent
where ${\cal C}^{\rm eq}(\omega)$ is the correlated part of the
equilibrium resolvent ${\cal R}^{\rm eq}(\omega)$.
The integral on the right-hand side of
equation (\ref{scr1.1}) has a structure similar to equation
(\ref{AXG.3}),
in that the adiabatic part of ${\cal R}^{\rm eq}$
makes no contribution after
decoupling of the intermediate wave-vector sums.

With equation (\ref{ccktb}), which gives
$\sum_{\alpha} {\cal C}^{\rm eq}_{\alpha \alpha'} = 0$,
summation over $\alpha$ in equations (\ref{scr1.1})
and (\ref{TF1}) produces

\begin{equation}
\gamma_{\rm C} = {{{\widetilde \Delta} N}\over {\Delta N}}
= \sum_{\alpha}
{{\delta \phi_{\alpha}}\over {\delta \mu}}
{{\Delta f^{\rm eq}_{\alpha}}\over {\Delta N}}
\label{TF5}
\end{equation}

\noindent
where ${{\widetilde \Delta} N} \equiv \sum_{\alpha}
{\widetilde \Delta} f^{\rm eq}_{\alpha}$.
In figure 1 we plot $\gamma_{\rm C}$ for the two-dimensional (2D)
electron gas forming the quantum-confined metallic channel in
a typical AlGaAs/InGaAs/GaAs heterojunction
\cite{vinter}.
The strong, self-consistent Coulomb contribution to the
energy levels in the quantum well gives a
pronounced dependence on carrier density. In turn,
this produces the negative feedback suppressing the
analogue to $\delta \phi_{\alpha}/\delta \mu$
within this system. The theory of
such systems is reviewed in the Appendix.

The resulting $\gamma_{\rm C}$
has immediate practical relevance to
heterojunction device engineering.
This factor modifies not only the
random thermal fluctuations, but also
the systematic variations of carrier
distribution in response
to a gate potential
\cite{vinter};
these determine the channel's current gain and
differential capacitance [see equation (\ref{Ap6})].
Through its direct effect on such quantities,
$\gamma_{\rm C}$ is a significant aspect of
high-performance transistor design.

We can compare equation (\ref{TF5}) with the variation in the
{\it contact potential} for transferring a conduction
electron from reservoir to sample.
The mean electrostatic potential per carrier is

\begin{equation}
u_{\rm cp}
= {1\over N} \sum_{\bf r} \Omega({\bf r})
V_0({\bf r}) {\langle f^{\rm eq}({\bf r}) \rangle}
= {1\over N} \sum_{\alpha} (\mu - \phi_{\alpha}) f^{\rm eq}_{\alpha}.
\label{scru}
\end{equation}

\noindent
Its variation is

\[
{{\delta u_{\rm cp}}\over {\delta \mu}}
= {1\over N} \sum_{\alpha}
{\left[
{\left(
1 - {{\delta \phi_{\alpha}}\over {\delta \mu}}
\right)} f^{\rm eq}_{\alpha}
+ (\mu - \phi_{\alpha} - u_{\rm cp})
{{{\widetilde \Delta} f^{\rm eq}_{\alpha}}\over {k_{\rm B}T}}
\right]},
\]

\noindent
which can be brought into the form

\begin{eqnarray}
{{\delta u_{\rm cp}}\over {\delta \mu}}
=&&
1 - \gamma_{\rm C}
+ \sum_{\bf r} \Omega({\bf r})
{{\delta V_0({\bf r})}\over {\delta \mu}}
{\left(
{{\langle f^{\rm eq}({\bf r}) \rangle}\over N}
- {{\langle \Delta f^{\rm eq}({\bf r}) \rangle}\over {\Delta N}}
\right)}
\cr
&&
+ {1\over k_{\rm B}T} \sum_{\bf r} \Omega({\bf r})
{\Bigl( V_0({\bf r}) - u_{\rm cp} \Bigr)}
{\langle {\widetilde \Delta} f^{\rm eq}({\bf r}) \rangle}.
\label{TF6}
\end{eqnarray}

\noindent
Evidently there is a close relationship between the
contact potential and $\gamma_{\rm C}$,
acting to renormalize the equilibrium thermal
fluctuations in an inhomogeneous sample.
For typical problems involving piecewise-uniform metallic structures,
the correction terms in equation (\ref{TF6}) will be small, so that
$\gamma_{\rm C} \to 1 - \delta u_{\rm cp}/\delta \mu$.

Figure 2 recapitulates the thermodynamic origin of $u_{\rm cp}$ in
a strongly metallic sample, placed between less-metallic leads.
This is a type of {\it p-n} junction formation.
Figure 3 contrasts the effect of complete versus partial
compensation of carrier fluctuations by the background response.
In 3(a) the response of the neutralizing background in the sample
totally compensates the carrier fluctuations
(as one assumes asymptotically in the leads). The contact
potential is immune to the presence of $\Delta f^{\rm eq}$,
which retains its bare, free-Fermi-liquid form.
In 3(b), the response of the background fails to screen out
completely the Coulomb potential generated by
$\Delta f^{\rm eq}$. The negative self-consistent feedback on
the fluctuations leads to their suppression.

The Boltzmann equation (\ref{scr1.1}) is solved by introducing
an equilibrium Coulomb screening operator
$\Gamma^{\rm eq}(\omega)$, whose inverse is

\begin{mathletters}
\label{scr1.2}
\begin{equation}
{\{ (\Gamma^{\rm eq})^{-1} \}}_{\alpha \alpha'}(\omega) 
= {\cal I}_{\alpha \alpha'}
+ {e\over \hbar}
\sum_{\beta}
{\cal C}^{\rm eq}_{\alpha \beta}(\omega)
{ {\partial f^{\rm eq}_{\beta}}
\over {\partial {\bf k_{\beta}}} } {\bbox \cdot}
{\widetilde {\bf E}}_{\rm C}({\bf r}_{\beta}, {\bf r'}),
\label{scr1.2a}
\end{equation}

\noindent
leading to the closed form

\begin{equation}
{\widetilde {\Delta}} f^{\rm eq}_{\alpha}
= \gamma_{\rm C} \sum_{\alpha'}
\Gamma^{\rm eq}_{\alpha \alpha'}(0) \Delta f^{\rm eq}_{\alpha'}.
\label{scr1.2b}
\end{equation}
\end{mathletters}

\noindent
In a non-uniform mesoscopic system, this is the counterpart to the
random-phase approximation for the static, screened fluctuations
in the standard rigid-jellium model of the uniform electron gas
\cite{pinoz}.

\subsection{Collision-mediated screening}

We now look beyond equilibrium, to the influence of dynamical
scattering. We start by defining the Coulomb
screening operator in an external driving field,
through the inverse

\begin{equation}
{\{ \Gamma^{-1} \}}_{\alpha \alpha'}(\omega) 
\buildrel \rm def \over =
{\cal I}_{\alpha \alpha'}
+ {e\over \hbar} \sum_{\beta}
{\cal C}_{\alpha \beta}(\omega)
{ {\partial f_{\beta}}
\over {\partial {\bf k_{\beta}}} } {\bbox \cdot}
{\widetilde {\bf E}}_{\rm C}({\bf r}_{\beta}, {\bf r'}).
\label{scr10.1}
\end{equation}

\noindent
The central role of this operator will become clear shortly. Here
we note that, while $\gamma_{\rm C}$ is collisionless,
$\Gamma$ exhibits the interplay of scattering and screening,
which is an exclusively non-equilibrium process.
Equation (\ref{ccktb}) means that $\Gamma$ has the property

\begin{equation}
\sum_{\alpha} \Gamma_{\alpha \alpha'}(\omega) = 1
{~~~~} {\rm for {~} all} {~} \alpha'.
\label{gprop}
\end{equation}

\noindent
This states that, while $\Gamma$ can shift fluctuation
strength at shorter scales within the structure, unlike
$\gamma_{\rm C}$ it cannot redistribute fluctuation
strength globally throughout the whole conductor.

As a basic illustration of collisional screening, in figure 4
we plot, for a simplified model
\cite{gdcond},
the collisional suppression factor $\gamma_{\rm coll}(q)$. This is
related to the real-space Fourier transform of $\Gamma$ via

\[
\gamma_{\rm coll}(q) = {2\over \Omega}
\sum_{\bf k} \Gamma_{{\bf k} {\bf k'}}({\bf q}, \omega = 0),
\]

\noindent
which turns out to be independent of the second wave vector ${\bf k'}$.
This partial average retains enough
structure to be a useful guide to the physics of $\Gamma$.
For technical details, see reference
\onlinecite{gdcond}.
At long wavelengths, $\gamma_{\rm coll}(q) \to 1$.
There is no suppression because the
collisional corrections are primarily local; the
effective interaction is fully screened by the leads
in the asymptotic limit, and accounts for
the inability of $\Gamma$ to renormalize
fluctuations across the structure. At very short wavelengths,
all screening (even full Thomas-Fermi) is irrelevant simply
because the bare interaction itself vanishes as $q^{-2} \ll 1$.
Again there is no suppression and $\gamma_{\rm coll}(q) \to 1$.

Collisional suppression is most effective at intermediate
distances. There it reflects the interplay of
material and geometrical parameters. One can expect a
rich variety of behaviours in this regard.

The formal significance of the non-equilibrium screening operator
is that it links the screening-free
propagators ${\cal R}$ and ${\cal C}$ to their
Coulomb-screened forms,
${\widetilde {\cal R}}$ and ${\widetilde {\cal C}}$.
In shortened notation, the linearized Boltzmann
equation with screening is

\begin{equation}
{\cal B}(\omega) \cdot {\widetilde {\cal R}}(\omega)
= {\cal I}
-{\left( {e\over \hbar}
{{\partial f}\over {\partial {\bf k}}}
{\bbox \cdot} {\widetilde {\bf E}}_{\rm C} \right)}
\cdot {\widetilde {\cal R}}(\omega).
\label{B1}
\end{equation}

\noindent
The same field ${\widetilde {\bf E}}_{\rm C}$ that is
defined in equation (\ref{scr0})
appears in both equilibrium and non-equilibrium situations.
That is because equations (\ref{poissoneq}) and ({\ref{poiss}),
combined, lead to $\delta {\bf E}({\bf r})/\delta f_{\alpha'}
= -{\widetilde {\bf E}}_{\rm C}({\bf r}, {\bf r'})$.
The latter result holds under the assumption that
$n_{\rm d}({\bf r})$ depends at most on $N$, whose
constancy away from equilibrium is enforced by the
overall neutrality of the conductor
through equation (\ref{gauss}).

Similarly, ${\widetilde \Delta} N$ is also invariant. This
follows from the zero norm of the total adiabatic propagator
${\widetilde G} \equiv \delta g/\delta f^{\rm eq}$,
when summed over the active volume.
We see that ${\widetilde G}$ is now the screened
version of the proper variational derivative $G$.
Exactly as with $G$, it is the global neutrality of $g$
[equation (\ref{gauss}) again] that guarantees the sum rule
for the steady-state fluctuation ${\widetilde \Delta} f$:

\[
\sum_{\alpha} {\widetilde \Delta} f_{\alpha}
= \sum_{\alpha}
{\Bigl( {\widetilde \Delta} f^{\rm eq}_{\alpha}
+ \sum_{\beta} {\widetilde G}_{\alpha \beta}
{\widetilde \Delta} f^{\rm eq}_{\beta} \Bigr)}
= {\widetilde \Delta} N.
\]

\noindent
This is the formal analogue to the
screening-free equation (\ref{DN}).
We do not elaborate the detailed form of
${\widetilde G}$, except to note that it is the zero-frequency
limit of the following screened operator
[compare equations (\ref{AXG.1}) and (\ref{AXR.2})]:

\[
{\widetilde {\cal G}}(\omega)
= {\widetilde {\cal R}}(\omega) \cdot 
({\widetilde {\cal R}^{\rm eq}})^{-1}(\omega) - {\cal I}.
\]

Equation (\ref{B1}) can be solved with ${\cal R}$:

\begin{equation}
{\widetilde {\cal R}}(\omega)
= {\cal R}(\omega) - {\cal C}(\omega) \cdot
{\left( {e\over \hbar}
{{\partial f}\over {\partial {\bf k}}}
{\bbox \cdot} {\widetilde {\bf E}}_{\rm C} \right)}
\cdot {\widetilde {\cal R}}(\omega)
= {\Gamma}(\omega) \cdot {\cal R}(\omega).
\label{B2}
\end{equation}

\noindent
In the screening term of the middle expression,
we omit the non-contributing adiabatic part of ${\cal R}$,
since ${\cal C}$ alone operates nontrivially within the integral.
In the rightmost expression, we have integrated the screening term
by applying equation (\ref{scr10.1}).
A match of residues in 
equation (\ref{B2}), at the pole $\omega = 0$,
provides the result [compare equation (\ref{AXR.3})]

\begin{equation}
{\widetilde \Delta} f_{\alpha}
= {{\widetilde \Delta} N}
\sum_{\alpha'} \Gamma_{\alpha \alpha'}(0)
{{\Delta f_{\alpha'}}\over {\Delta N}}
= \gamma_{\rm C}
\sum_{\alpha'} \Gamma_{\alpha \alpha'}(0)
\Delta f_{\alpha'}.
\label{B2.1}
\end{equation}

\noindent
This is the first of two key equations of this section.
It affords an exact and explicit formula
for the steady-state fluctuations,
extending the random-phase screened equation (\ref{scr1.2b})
to the driven system.
Above all, it establishes that the non-equilibrium,
Coulomb-screened thermal fluctuations {\it must scale linearly
with the equilibrium renormalization} $\gamma_{\rm C}$.
This, in turn, means that the current auto-correlation will
also be proportional to $\gamma_{\rm C}$.
We emphasize that this scaling principle
is the strict outcome of asymptotic equilibrium and
charge balance in the macroscopic leads.

\subsection{Dynamics and the current auto-correlation}

We now obtain the dynamics with screening,
given by the correlated part ${\widetilde {\cal C}}$
of the resolvent ${\widetilde {\cal R}}$.
Equation (\ref{B2}) is equivalent to

\begin{equation}
{\widetilde {\cal C}}(\omega)
- {{\cal I}\over {i(\omega + i0^+)}} \cdot
{ {{\widetilde \Delta} f}\over {{\widetilde \Delta} N} }
= \Gamma(\omega) \cdot
{\left(
{\cal C}(\omega)
- {{\cal I}\over {i(\omega + i0^+)}} \cdot
{{\Delta f}\over {\Delta N}}
\right)},
\label{C1}
\end{equation}

\noindent
which can be rearranged, first with help from Eq. (\ref{B2.1}),
then by taking the product with ${\widetilde \Delta} f$ 
on both sides and invoking
the screened form of equation (\ref{cckta}),
${\widetilde {\cal C}}(\omega) \cdot {\widetilde \Delta} f = 0$:

\begin{eqnarray}
{\widetilde {\cal C}}(\omega)
=&&
\Gamma(\omega) \cdot {\cal C}(\omega)
- {{\Gamma(\omega) - \Gamma(0)}\over {i(\omega + i0^+)}} \cdot
{{\Delta f}\over {\Delta N}}
\cr
{\left. \right.} \cr
=&&
\Gamma(\omega) \cdot {\cal C}(\omega) \cdot
{\left( 
{\cal I} - { {{\widetilde \Delta} f}\over {{\widetilde \Delta} N} }
\right)}.
\label{C2}
\end{eqnarray}

\noindent
This is the second key equation of the model.
An alternative form, for later use, is

\begin{equation}
\Gamma(\omega) \cdot {\cal C}(\omega)
= {\widetilde {\cal C}}(\omega) \cdot
{\left( 
{\cal I} - { {\Delta f}\over {\Delta N} }
\right)}.
\label{C3}
\end{equation}

With equation (\ref{C2}) we gain the last component
of the microscopic velocity auto-correlation, fully screened.
In complete analogy with the screening-free correlation
\cite{gdi},
its zero-frequency form is

\begin{equation}
{ \langle\!\langle {\bf v} {\bf v'}
{ {\widetilde \Delta} {\rm f}^{(2)} }({\bf r}, {\bf r'}; 0) 
\rangle\!\rangle}_{\rm c}'
\buildrel \rm def \over =
{1\over \Omega({\bf r})} {1\over \Omega({\bf r'})}
\sum_{{\bf k}, s} \sum_{{\bf k'}, s'}
{\bf v}_{{\bf k} s}
{\rm {\widetilde {\cal C}}}_{\alpha \alpha'}(0)
{\bf v}_{{\bf k'} s'}
{\widetilde \Delta} f_{\alpha'},
\label{scr16}
\end{equation}

\noindent
where we keep the correlated part of
${\widetilde \Delta} {\rm f}^{(2)}_{\alpha \alpha'}(\omega)
= {\widetilde {\cal R}}_{\alpha \alpha'}(\omega)
{\widetilde \Delta} f_{\alpha'}$.
At finite frequency one must add the displacement-current
contribution to the fluctuations. The velocity
is replaced with the non-local operator

\begin{equation}
{\bf u}_{{\bf k} s}({\bf r}, {\bf r''}; \omega)
\equiv 
{{\delta_{ss''}\delta_{{\bf r} {\bf r''}}}\over \Omega({\bf r''})}
{\bf v}_{{\bf k''} s''}
- {{i\omega \epsilon}\over {4\pi e}}
{\bf E}_{\rm C}({\bf r} - {\bf r''})
\label{displ}
\end{equation}

\noindent
where $e{\bf E}_{\rm C}({\bf r} - {\bf r''})
\equiv -\partial V_{\rm C}({\bf r} - {\bf r''})/\partial {\bf r}$
is the bare Coulomb force, coupling to the dynamically
changing carrier density.
The introduction of displacement currents means that
two additional, intermediate sums over region $\Omega$
must be incorporated within the expectation
$\langle\!\langle
{\rm Re} {\{\bf u} {\widetilde \Delta} {\rm f}^{(2)} {\bf u'}^*\}
\rangle\!\rangle'_{\rm c}$. At $\omega = 0$ it
reverts to equation (\ref{scr16}).

As with the screening-free version
\cite{gdi},
equation (\ref{scr16})
conveys the physics of the spontaneous electron-hole excitation
${\widetilde \Delta}{\rm f}^{(2)}$ in
the metallic conduction band. Its average thermal strength is
${\widetilde \Delta} f$, giving the initial flux contribution
${\bf v'} {\widetilde \Delta} f$.
Once it is spontaneously excited out of the steady-state background,
the thermal pair excitation evolves semi-classically
{\it in keeping with the full Boltzmann equation of motion},
equation (\ref{B1}), characterized by ${\widetilde {\cal C}}(\omega)$.
The final weighting by ${\bf v}$ sets up the auto-correlation
of the microscopic flux.

The whole process describes the non-equilibrium development of the
electron-hole fluctuations, starting out
as elementary excitations out of the steady state of the electron liquid
and relaxing dynamically back to the steady state.
There is much more in equation (\ref{scr16})
than in its unscreened analogue
\cite{gdi}.

\section{Fluctuation-dissipation theorem}

The fluctuation-dissipation theorem (FDT) is the prime relation of
linear low-field transport, and its derivation within a given model
gives first-hand evidence of that model's internal consistency.
We prove that the FDT is satisfied within our description
of non-uniform mesoscopic Coulomb systems; since our approach
is a kinetic-theoretical one our derivation of the mesoscopic
FDT is closer to van Kampen's programme for response theory
\cite{nvk}
than it is to Kubo's
\cite{kubo}.
The former viewpoint is better suited to non-perturbative
calculations in the high-field regime, which is
important for applications
to practical devices where one must go beyond the FDT.
After establishing the theorem,
we look at its experimental implications.

To make contact with measurable properties such as the conductance and
the thermal noise-power spectrum,
both the single-particle distribution $g$ and the two-particle
correlation $\langle\!\langle {\bf v} {\bf v'} {\widetilde \Delta}
{\rm f}^{(2)}(0) \rangle\!\rangle'_{\rm c}$ should be
related directly to the externally applied field ${\bf E}_{\rm ext}$.
(For convenience we take a uniform dielectric constant
$\epsilon$.)  Recalling the Poisson
equation (\ref{poiss}) for the induced field ${\bf E}_{\rm ind}$
as a functional of $g$, the kinetic equation (\ref{AX3}) for
$g$ itself can be transformed:

\begin{eqnarray}
{\cal B}(0) \cdot g
=&& {e\over \hbar}
( {\bf E}_{\rm ext} + {\bf E}_{\rm ind}[g] )
{\bbox \cdot} {{\partial f^{\rm eq}}\over {\partial \bf k}}
- {\cal W}^{(2)}[g]
\cr
{\left. \right.} \cr
=&& - {e\over k_{\rm B}T} {\bf E}_{\rm ext}
{\bbox \cdot} {\bf v} {~}\Delta f^{\rm eq}
- {e\over \hbar} {{\partial f^{\rm eq}}\over {\partial \bf k}}
{\bbox \cdot} ({\bf E}_{\rm C} \cdot g)
- {\cal W}^{(2)}[g].
\label{AX3.1}
\end{eqnarray}

\noindent
In the leading term on the right-hand side of the second line,
the identity $\partial f^{\rm eq}/\partial {\bf k}
= -(\hbar {\bf v}_{\bf k}/k_{\rm B}T) \Delta f^{\rm eq}$
has been used 
\cite{jrho}.
In the middle right-hand term, the bare Coulomb force
arises from the integral form of the Poisson equation
for ${\bf E}_{\rm ind}[g]$.

We now make the first of two simplifying assumptions:
let the neutralizing background of the conductor stay fixed
(rigid jellium), so that it does not cancel the self-consistent
response of the fluctuations.
Then the effective field ${\widetilde {\bf E}}_{\rm C}$
becomes identical with ${\bf E}_{\rm C}$,
and we may integrate the Boltzmann equation (\ref{AX3.1})
by invoking the screening operator $\Gamma$:

\begin{eqnarray}
g
=&& 
-{\cal C}(0) \cdot
{\left( {e\over k_{\rm B}T} {\bf E}_{\rm ext}
{\bbox \cdot} {\bf v} {~}\Delta f^{\rm eq}
+ {e\over \hbar}
{{\partial (f - g)}\over {\partial \bf k}}
{\bbox \cdot} ({\bf E}_{\rm C} \cdot g)
+ {\cal W}^{(2)}[g]
\right)}
\cr
{\left. \right.} \cr
\equiv&&
-{e\over k_{\rm B}T} \Gamma(0) \cdot {\cal C}(0) \cdot
({\bf E}_{\rm ext} {\bbox \cdot} {\bf v} {~}\Delta f^{\rm eq})
+ h
\label{AX3.2}
\end{eqnarray}

\noindent
where 

\[
h = \Gamma(0) \cdot {\cal C}(0) \cdot 
{\left(
{e\over \hbar} {{\partial g}\over {\partial \bf k}}
{\bbox \cdot} ({\bf E}_{\rm C} \cdot g)
- {\cal W}^{(2)}[g]
\right)}
\]

\noindent
is the remnant non-linear term, of order $g^2$.
Next we construct the total resistive power dissipation $P(\Omega)$
over the conducting region $\Omega$. The local current density
is ${\bf J}({\bf r}) = -e{\langle {\bf v} g({\bf r}) \rangle}$,
so that

\begin{equation}
P(\Omega)  = \sum_{\bf r} \Omega({\bf r})
{\widetilde {\bf E}}({\bf r}) {\bbox \cdot}
{\bf J}({\bf r})
= -e \sum_{\alpha} ({\bf E}_{\rm ext}
{\bbox \cdot} {\bf v})_{\alpha} g_{\alpha}.
\label{FDT1}
\end{equation}

\noindent
There is no contribution to $P$ from the induced field
${\bf E}_{\rm ind}({\bf r})$.
First, the total current $I$ is conserved since the flux density,
integrated over any directed surface bisecting
the conductor, is constant.
Second, the internal Coulomb forces are conservative so
that the difference in {\it induced} potential between
two points on any circuit path through $\Omega$,
one point deep in the source and the other in the drain,
must vanish. Thus the current may be factored out of the
right-hand side of equation (\ref{FDT1}) to leave just a
line integral for the total electromotive potential $V$
and resulting in the canonical dissipation formula
$P(\Omega) = IV$. The microscopic proof of this fundamental
transport theorem, under the most general mesoscopic conditions,
was given recently by Magnus and Schoenmaker
\cite{wims};
it is absolutely essential to the FDT.

Note that $P(\Omega)$ is finite and calculable if and only if
$\Omega$ is indeed a {\it bounded} region. Once again the
hypotheses of equilibrium and charge balance in the leads
are indispensable.

Our other simplifying assumption is that the applied field
${\bf E}_{\rm ext}$ is uniform over $\Omega$,
with no internal sources. We take the field to be
constant along the $x$-axis, whose sense is from source to drain
\cite{emf}.
Let it be $-V/L$ over sample length $L$.
Equations (\ref{AX3.2}) and (\ref{FDT1}) combined lead to

\begin{eqnarray}
P(\Omega)
=&&
\sum_{\alpha} (-e{\bf E}_{\rm ext} {\bbox \cdot} {\bf v})_{\alpha}
{\left(
{1\over k_{\rm B}T}
\sum_{\alpha'} [\Gamma(0) \cdot {\cal C}(0)]_{\alpha \alpha'}
(-e{\bf E}_{\rm ext} {\bbox \cdot} {\bf v})_{\alpha'}
\Delta f^{\rm eq}_{\alpha'} + h_{\alpha}
\right)}
\cr
{\left. \right.} \cr
\equiv&&
{1\over k_{\rm B}T}{\left( {eV\over L} \right)}^2
\sum_{\alpha \alpha'}
v_x [ \Gamma(0) \cdot {\cal C}(0) ]_{\alpha \alpha'}
v'_x \Delta f^{\rm eq}_{\alpha'} + P_h(\Omega).
\label{FDT2}
\end{eqnarray}

\noindent
The contribution $P_h(\Omega)$ is of order $Vg^2$;
this non-linear complement is negligible in the weak-field limit.
The first line of equation (\ref{FDT2}) highlights the fact that the
mean dissipative power is itself an {\it auto-correlation function for
the power density} ${\bf E}_{\rm ext} {\bbox \cdot} {\bf J}$.
This is the heart of the FDT, in the context of electron transport.

To make a final connection with device parameters,
we express the current phenomenologically as the usual
relation $I \equiv GV$, where $G$ is the conductance
(not to be confused with earlier notation for the
adiabatic propagator). Since $P = IV$, we have

\begin{equation}
G = P(\Omega)/V^2 \equiv {\cal S}(0)/4k_{\rm B}T
\label{FDT3}
\end{equation}

\noindent
in terms of the low-frequency current-noise spectral density
\cite{ggk,nougier}

\begin{equation}
{\cal S}(0)
= 4 \sum_{\alpha \alpha'}
( -{ev_x}/L ) [\Gamma(0) \cdot {\cal C}(0)]_{\alpha \alpha'}
( -{ev'_x}/L ) \Delta f^{\rm eq}_{\alpha'}
+ {\cal S}_h(0)
\label{FDT4}
\end{equation}

\noindent
where the non-linear correction
${\cal S}_h(0) = 4 k_{\rm B}T P_h(\Omega)/V^2$ is well-defined;
close to equilibrium, only the first right-hand sum survives.
Note, however, that equation (\ref{FDT4}) holds for arbitrary
driving fields.

Equations (\ref{FDT3}) and (\ref{FDT4}) recover
the Johnson-Nyquist formula and establish the
fluctuation-dissipation theorem for a mesoscopic system,
complete with Coulomb effects.
The theorem relates the empirical Joule-heating rate
over the sample, determined by $G$,
to the equilibrium noise power determined by ${\cal S}(0)$.
We remark that {\it collisional screening}, through $\Gamma(0)$
in equation (\ref{FDT4}), will be an important modifier of
the conductance in short non-uniform samples.

We must now address the role of
the {\it microscopic} current auto-correlation,
equation (\ref{scr16}). Let

\begin{eqnarray}
{\widetilde {\cal S}}(0)
=&&
4 \sum_{\bf r} \Omega({\bf r}) \sum_{\bf r'} \Omega({\bf r'})
{ \langle\!\langle (-e v_x/L) (-e v'_x/L)
{ {\widetilde \Delta} {\rm f}^{(2)} }({\bf r}, {\bf r'}; 0) 
\rangle\!\rangle}_{\rm c}'
\cr
{\left. \right.} \cr
=&&
4 {e^2\over L^2} \sum_{\alpha \alpha'}
v_x {\widetilde {\cal C}}(0)_{\alpha \alpha'}
v'_x ( {\widetilde \Delta} f^{\rm eq} +
{\widetilde {\cal G}}(0) \cdot {\widetilde \Delta} f^{\rm eq} )_{\alpha'}
\cr
{\left. \right.} \cr
\equiv&&
4 {e^2\over L^2} \sum_{\alpha \alpha'}
v_x {\widetilde {\cal C}}(0)_{\alpha \alpha'}
v'_x {\widetilde \Delta} f^{\rm eq}_{\alpha'}
+ {\widetilde {\cal S}}_g(0).
\label{FDT5}
\end{eqnarray}

\noindent
This fully non-equilibrium spectral density
includes the well-defined excess
contribution ${\widetilde {\cal S}}_g(0) \propto
{\widetilde {\cal G}}(0) \cdot {\widetilde {\Delta}} f^{\rm eq}$;
in the weak-field limit we may drop this,
since it vanishes with $g$.
With equation (\ref{C3}) we transform (\ref{FDT4}) into

\begin{eqnarray}
{\cal S}(0)
=&&
4 {e^2\over L^2} \sum_{\alpha \alpha'}
v_x {\left[
{\widetilde {\cal C}}(0) \cdot
{\left( {\cal I}
- {{\Delta f}\over {\Delta N}}
\right)} \right]}_{\alpha \alpha'}
v'_x
\Delta f^{\rm eq}_{\alpha'}
\cr
{\left. \right.} \cr
=&&
4 {e^2\over L^2} \sum_{\alpha \alpha'}
v_x {\widetilde {\cal C}}(0)_{\alpha \alpha'}
v'_x \Delta f^{\rm eq}_{\alpha'}.
\label{FDT6}
\end{eqnarray}

\noindent
The sum for $(v_x {\widetilde {\cal C}}(0) \cdot
\Delta f) \cdot (v'_x \Delta f^{\rm eq})$ vanishes because
${\langle v_x {\widetilde {\cal C}}(0) \cdot \Delta f \rangle}$
decouples
from ${\langle v'_x \Delta f^{\rm eq} \rangle}' = 0$.
By adding and subtracting
${\widetilde {\Delta}} f^{\rm eq}_{\alpha'}/\gamma_{\rm C}$
from $\Delta f^{\rm eq}_{\alpha'}$ on the right-hand side
of equation (\ref{FDT6}) we obtain the strictly low-field
relation

\begin{mathletters}
\label{FDT7}

\begin{equation}
{\cal S}(0)
= {1\over \gamma_{\rm C}}
{\left(
{\widetilde {\cal S}}(0)
- 4 {e^2\over L^2} \sum_{\alpha \alpha'}
v_x {\widetilde {\cal C}}(0)_{\alpha \alpha'}
v'_x ( {\widetilde \Delta} f^{\rm eq}_{\alpha'}
- \gamma_{\rm C} \Delta f^{\rm eq}_{\alpha'} )
\right)}.
\label{FDT7a}
\end{equation}

\noindent
The link between ${\cal S}(0)$ and ${\widetilde {\cal S}}(0)$
is most transparent when the correction term is small.
This will be so under
the same conditions in which the relation $\gamma_{\rm C}
\approx 1 - \delta u_{\rm cp}/\delta \mu$ applies:
the non-uniformity of the structure must reside
mostly in the interfacial zones, which will be
short compared to $L$ so that they do not overlap. Then
${\widetilde \Delta} f^{\rm eq} \to \gamma_{\rm C} \Delta f^{\rm eq}$
and

\begin{equation}
{\cal S}(0) \to {\widetilde {\cal S}}(0)/\gamma_{\rm C}.
\label{FDT7b}
\end{equation}
\end{mathletters}

Before discussing the experimental consequences of this section,
we make an important observation. In comparing the noise
spectral densities of equations (\ref{FDT4}) and (\ref{FDT5}),
the crucial point is not the striking difference in scale,
$\gamma_{\rm C}$,
but the fact that {\it they evolve as completely different
mathematical structures}. In the high-field limit one or both of
the purely non-linear contributions, ${\cal S}_h(0)$
and ${\widetilde {\cal S}}_g(0)$, will not be negligible.

Even when the sample and its reservoirs are of identical material,
with no contact-potential effects, the correlation functions
which generate ${\cal S}(0)$ and ${\widetilde {\cal S}}(0)$
do {\it not} describe the same underlying physics.
The former deals with resistive dissipation of electrical energy,
while the latter deals with the dynamical spreading of
correlations in real space and momentum space.
These are quite distinct (though clearly connected)
dynamical processes.
Their separateness should manifest itself in the
non-equilibrium current noise, and nowhere more strongly
than in a {\it Coulomb-correlated system}.

\section{Experimental implications}

\subsection{Equilibrium versus high fields}

The fluctuation-dissipation theorem asserts
the necessary equivalence of ${\cal S}(0)$ and
${\widetilde {\cal S}}(0)$ in the weak-field limit
(leaving aside, for now, the contact potential);
clearly, the FDT can say nothing about the non-equilibrium region.
In GD we discussed the main physical distinction between
resistive and non-linear fluctuations
\cite{gdi,gdcond}:
the existence of {\it non-dissipative} excess
noise entering via ${\widetilde {\cal S}}_g(0)$.

Our model demands that, for degenerate conductors,
hot-electron noise must remain proportional to ambient temperature.
Here we recall the indirect but strong evidence
in favour of overall $T$-scaling of thermal noise,
in the observations of Liefrink {\it et al}.
for 2D mesoscopic wires
\cite{liefr}.
They anticipated -- yet failed to detect -- a large
supra-thermal signature from electron-electron scattering
(two-body collisional effects are particularly enhanced in 2D).
According to standard estimates,
that signature {\it should} have scaled with a
hot-carrier temperature $T_{\rm e} \gg T$.
Remarkably enough, no such contribution was seen.

The experimental reality and accessibility of
non-equilibrium, non-dissipative fluctuations is not at issue.
Shot noise, proportional to the current, is the foremost
example of a stochastic process not associated with
dissipation (else equation (\ref{FDT3}) would entail
a linear current-dependent correction in $G$; such a
correction is not observed with shot noise).
As far as noise that is generated strictly thermally is concerned,
there are two major experimental questions, as yet unanswered:

\begin{itemize}
\item Where is the onset of hot-electron behaviour in mesoscopic noise?
\item What is its temperature dependence in the degenerate regime?
\end{itemize}

\noindent
Even without heavy calculation
\cite{gdcond,mbix}
our formalism predicts
the growth of hot-electron noise, scaling with $T$.
At the same time, theoretical work by Shimizu and Ueda
\cite{shimi}
and by Liu, Eastman and Yamamoto
\cite{liu}
suggests that the $T$-independent shot noise should be attenuated
as inelastic phonon scattering  begins to outweigh
elastic impurity scattering at higher fields.

\subsection{Screening}

We come to the physics of contact-potential screening. Our equation
(\ref{FDT5}) unambiguously requires suppression of the intrinsic
current fluctuations for an inhomogeneous sample, relative to the
resistive Johnson-Nyquist spectral density for the same system.
We emphasize that the correlation ${\widetilde {\cal S}}(0)$
follows directly from the exact solution to the linearized,
fully self-consistent, Boltzmann equation.
Since ${\widetilde {\cal S}}(0)$ 
dictates the fluctuations underlying the system,
it is in no sense an artificial construct;
it conveys crucial physical information.

At low fields, the factoring-out of $\gamma_{\rm C}$ in equation
(\ref{FDT7}) ``saves the phenomenon'' of the FDT
for the power density by removing all trace
of contact-potential effects from the microscopic spectral density,
which then recovers the conventional Johnson-Nyquist form
\cite{class}.
An identical rescaling procedure, preserving the FDT,
has been discussed by Kittel
\cite{kittel}
for a dissipative system in which
(non-dissipative) active feedback works to reduce the fluctuations.

At high fields, the story takes a very different turn. The specific
functional form of ${\widetilde {\cal S}}_g(0)$
means that one should see suppression at least
in the {\it hot-electron noise}
\cite{gdi}.
This should survive any of the calibration protocols commonly used
to separate the sought-after noise signal from the unwanted background.

In the quasi-equilibrium limit, the choice of calibration strategy
is crucial to the acquisition of data on screening in noise.
For example, the equilibrium term was subtracted from the early 2D
measurements of Liefrink {\it et al}. for shot noise
\cite{liefr}.
Screening-suppressed current noise, if present in the quiescent state,
would have been lost from their low-field signal. In other cases,
such as the Steinbach {\it et al}. shot-noise experiment in
three-dimensional (3D) silver wires
\cite{smd},
the equilibrium
noise floor is used as an absolute check on the measurement scale.
However, since $\gamma_C = 1$ in homogeneous 3D metallic structures,
there is no screening to detect. Therefore, a combination of 2D wire
samples (for screening) and non-subtractive calibration
(to keep the equilibrium current correlations) would be ideal.

\subsection{Experiments suggested}

Testing the suppression theory calls for a specific structure in
which screening is not only present, but easily controlled.
Such an arrangement was suggested in reference
\onlinecite{upon}.
Here we show it in figure 5, with more detail.
It relies on a back-gated diffusive wire,
fabricated on a heterojunction
quantum-well substrate to provide the conductive medium:
a uniform, two-dimensional and tunable metallic electron gas.
The theory of $\gamma_{\rm C}$ in quantum-confined 2D
electron gases is reviewed in the Appendix (see also figure 1).
Although $\gamma_{\rm C}$ in 2D differs mathematically
from its 3D form (section 2.2 above), their Coulomb origins and
physical consequences are the same.

In a channel that is uniform in the plane of conduction,
the non-linear term ${\cal S}_h(0)$ of
equation (\ref{FDT4}) contains
no screening correction. It depends only on the residual collision
kernel ${\cal W}^{(2)}$ [within $h$ in equation (\ref{AX3.2})],
which is zero for pure elastic scattering
(although inelastic phonon and inter-band scattering
will revive it for large enough $V$).
Homogeneity also means that any correction terms, such as the one
on the right-hand side of equation (\ref{FDT7}), are small
\cite{caveat}.
Thus, of the two disparate non-linear corrections
${\cal S}_h(0)$ and ${\widetilde {\cal S}}_g(0)$, only the
latter is significant and should appear explicitly as
part of the thermal current-current correlation signal.
Its characteristic experimental signatures should be
\smallskip

(a) linear dependence of excess noise on bath temperature $T$, and
\smallskip

(b) linear dependence of excess noise on the suppression factor
$\gamma_{\rm C}$.
\smallskip

\noindent
Any controlled modulation of $T$ and $\gamma_{\rm C}$ should
be mirrored in the non-equilibrium noise spectrum.
Previously we have reasoned
\cite{gdcond,upon}
that (b), just like (a), cannot apply to shot noise; its
contribution to the total dissipationless noise
should depend neither on $T$ nor on $\gamma_{\rm C}$.
This means that thermal and shot noise ought to be clearly
separable from each other,
according to their different response to a gate-bias
potential. In this way each becomes individually accessible
for study within the same experiment. A method for the direct
resolution of the two kinds of noise has not been available up to now.

Such predictions, if correct, should shed light on the
behaviour of current
fluctuations beyond linear response. Ultimately, non-linear processes
will be the determinants of device performance in technological
applications of mesoscopic electronics. Any technique that advanced
the development of non-equilibrium physics would be of value.

\section{Summary}

Our aim has been to set up a systematic account of current fluctuations
in inhomogeneous metallic conductors, at mesoscopic scales. Its vehicle
is the semi-classical kinetic equation, supplemented with the theory of
charged Fermi liquids. These are the definitive, universally recognized
tools of transport analysis.

Our conclusion is that Coulomb screening exerts a marked influence on the
fluctuations of degenerate carriers in conductors that are highly
non-uniform. The resulting screening-induced suppression of current noise
should be observable in several different ways. An appropriate system
for such measurements is the two-dimensional electron gas.

This paper's complexity makes it useful to retrace our path.

\begin{itemize}
\item
We started with the Boltzmann description of non-equilibrium transport
and fluctuations for a mesoscopic conductor, connected to macroscopic
current leads. The boundary conditions in the leads, namely {\it local
neutrality} and {\it local equilibrium} at all times, govern the form
of the non-equilibrium fluctuations within the driven conductor.
The absolutely cardinal role of these constraints cannot be
stressed enough. In particular they ensure that the thermal noise of a
degenerate system, even away from equilibrium, always retains
its characteristic equilibrium feature: proportionality
to the ambient temperature.

\item
Next, we extended the kinetic description to cover the internal fields'
self-consistent response to electron-hole pair excitations, which make
up the fluctuations. We showed how inhomogeneity between the conductor
and its electron reservoirs sets the scene for overall suppression of
thermal current noise. This is produced by negative feedback from
fluctuations of the contact potential between sample and reservoirs.
A separate, non-equilibrium Coulomb process arises in association with
scattering. This type of screening acts at shorter range, and has
direct influence on the conductance if the channel is small enough.
Collisional screening, unlike contact-potential screening,
cannot alter the {\it overall} scale of noise throughout the channel.

\item
We then proved the fluctuation-dissipation theorem within
our kinetic theory of inhomogeneous mesoscopic conductors.
We pointed out the sharp physical distinction
between correlations in the {\it power density},
which are fundamentally
tied to resistive dissipation, and spontaneous fluctuations
in the {\it current density}, which are explicitly suppressed by
contact-potential screening and which develop a non-dissipative
excess component out of equilibrium. Nevertheless, at low fields
there exists a simple quantitative relationship
between power-density and
current-density fluctuations. This is required by the thermodynamics
of the fluctuation-dissipation theorem.

\item
Last, we proposed an experimental arrangement to probe both
Coulomb-screening and non-dissipative features of current noise.
This is best done in a two-dimensional channel, fabricated on a back-gated
heterojunction for uniform control over carrier density and the
amount of screening. Care should be taken with low-field calibration
of current noise in the two-dimensional wire, to retain correlations
in the quiescent state.
\end{itemize}

We have aimed to keep faith with basic principles of kinetic and
electron-gas theory, albeit at a price in immediate formulaic appeal.
One's concern should be primarily for physical substance,
less for cosmetics.
Still, our own philosophy may continue to be viewed, by some,
as mesoscopic revisionism in the face of a prevailing
semi-classical doctrine
\cite{kogancpu}.
In reality, the complete opposite is true.
The present description sticks manifestly
to the letter of the kinetic orthodoxy.
It rejects all generic assumptions that are logically unnecessary
to transport and demonstrably inconsistent
with the physics of charged Fermi liquids
\cite{ithaca,upon}.
Its outcomes are entirely conservative
in spirit and guaranteed to be fully conserving in practice.
Our experimental predictions are specific and verifiable. 

\section*{Acknowledgments}

We thank J. Wiggins and R. Smith for valuable suggestions on the
design and fabrication of 2D structures for non-linear noise
experiments, and J. Leahy for help with illustrations.

\appendix
\section{Screening in self-confined quantum channels}

For convenience, we describe the origin of screening suppression
in heterojunction quantum-well channels. Our treatment presupposes
the practical solution of the bound-state
Schr\"odinger problem in these structures.
Details are in the literature
\cite{vinter,bastard}.

Assume that the bound-state energy levels
$\{\varepsilon_i(n_s)\}_i$ have been computed for an ensemble of
carriers, confined in the quantum well with uniform 2D density $n_s$.
Quantization is transverse to the plane of free motion
for the carriers. The conducting states form discrete sub-bands
whose energy thresholds are $\varepsilon_i(n_s)$.
Since the Hamiltonian includes the large self-consistent
potential from mutual Coulomb repulsion of the localized electrons,
the levels naturally depend on $n_s$. The associated equilibrium
distribution in sub-band $i$, according to equation (\ref{AX1.2}), is

\begin{equation}
f^{\rm eq}_{i, {\bf k}} ~=~
{\left[
1 + \exp \!
 {\left(
   { {\varepsilon_{\bf k} + \varepsilon_i(n_s) - \mu}\over k_{B}T }
  \right)}
\right]}^{-1}.
\label{Ap1}
\end{equation}

The density in device area $\Omega$
fixes $\mu$ implicitly through

\begin{equation}
n_s
= {2\over \Omega} \sum_{i, {\bf k}} f^{\rm eq}_{i, {\bf k}}
= n^*_s \sum_i \ln{\{ 1 + \exp[(\mu - \varepsilon_i(n_s))/k_{\rm B}T] \}}
\label{Ap2}
\end{equation}

\noindent
where $n^*_s \equiv m^*k_{\rm B}T/\pi\hbar^2$ is the natural
2D density scale ($m^*$ is the effective mass).
The density of {\it free} electron-hole pair fluctuations is
straightforward to calculate:

\begin{equation}
\Delta n_s
= k_{\rm B}T {{\partial n_s}\over {\partial \mu}}
= n^*_s \sum_i f^{\rm eq}_{i, {\bf 0}}. 
\label{Ap3}
\end{equation}

\noindent
The dimensionless degeneracy factor $\Delta n_s/n_s$ is shown
in figure 1. The smaller it is, the greater the degeneracy of the
carriers; the closer to unity, the more classically they behave.

Equation (\ref{Ap3}) ignores the self-consistency
of the bound-state energies. When this is included
one obtains the density of {\it Coulomb-correlated} pair fluctuations

\begin{equation}
{\widetilde \Delta} n_s
= k_{\rm B}T {{\delta n_s}\over {\delta \mu}}
= n^*_s \sum_i
{\left( 1 - {{\delta \varepsilon_i}\over {\delta \mu}} \right)}
f^{\rm eq}_{i, {\bf 0}}. 
\label{Ap4}
\end{equation}

\noindent
Similarly, the fluctuation in $f^{\rm eq}_{i, {\bf k}}$ is

\[
{\widetilde \Delta} f^{\rm eq}_{i, {\bf k}}
= {\left( 1 - {{\delta \varepsilon_i}\over {\delta \mu}} \right)}
\Delta f^{\rm eq}_{i, {\bf k}}.
\]

\noindent
The factor in parentheses plays the role of
$\delta \phi_{\alpha}/\delta \mu$ in our description of
3D mesoscopic screening. We have a system that is uniform in
the plane of conduction but highly inhomogeneous in the
transverse direction. Its reduced symmetry, which
is not immediately apparent at the level of one-body transport,
is nevertheless a powerful modifier of the fluctuations
within the plane.

Equation (\ref{Ap4}) can be closed for ${\widetilde \Delta n_s}$
by writing

\[
{{\delta \varepsilon_i}\over {\delta \mu}}
= {{d \varepsilon_i}\over {d n_s}}
{{{\widetilde \Delta} n_s}\over k_{\rm B}T}
\]

\noindent
so that a term proportional to ${\widetilde \Delta} n_s$ can
be transferred from right to left in the equation. The result is

\begin{equation}
{\widetilde \Delta} n_s
=
{\displaystyle
{ {n^*_s {\sum}_i f^{\rm eq}_{i, {\bf 0}}}
}
\over
{\displaystyle
{1 + {n^*_s\over k_{\rm B}T} {\sum}_i
{{d \varepsilon_i}\over {d n_s}} f^{\rm eq}_{i, {\bf 0}}
} }
} {~}\equiv{~} \gamma_{\rm C} \Delta n_s.
\label{Ap5}
\end{equation}

\noindent
This is the 2D screening-suppression factor.
With allowance for dimensionality,
all of the Coulomb screening theory in sections 2 and 3 of the
main text is rigorously applicable to the 2D case.
In practice, most carriers in a degenerate 2D channel
occupy the ground-state sub-band, $i = 0$. In that case
we have $\gamma_{\rm C} \approx 1 - \delta \varepsilon_0/\delta \mu$,
and ${\widetilde \Delta} f^{\rm eq}_{0, {\bf k}}
\approx \gamma_{\rm C} \Delta f^{\rm eq}_{0,{\bf k}}$.

Finally, we show that the same suppression factor enters
directly into the measurable parameters of an active,
gated structure built on the heterojunction
\cite{vinter}.
We consider the differential capacitance
of the carrier population under a biasing gate (of area $\Omega$).
A change in bias potential $V_{\rm g}$ at the gate, which
overlies the modulation-doped layer above the channel, alters the profile
$n_{\rm d}(z)$ of the ionized donors in the transverse
$z$ direction. In turn this alters $n_s$.
Solution of the Poisson donor-depletion problem provides the
variational derivative $\delta \mu/\delta V_{\rm g}$. The channel's
contribution to $C_{\rm gs}$, the differential capacitance,
is the change in total charge $Q = e\Omega n_s$ in the channel. Thus

\begin{equation}
C_{\rm gs}
\sim -{{\delta Q}\over {\delta V_{\rm g}}}
= e\Omega {{\delta n_s}\over {\delta \mu}}
{\left| {{\delta \mu}\over {\delta V_{\rm g}}} \right|}
= \gamma_{\rm C}
{\left( e\Omega {{\partial n_s}\over {\partial \mu}} 
{\left| {{\delta \mu}\over {\delta V_{\rm g}}} \right|}
\right)}.
\label{Ap6}
\end{equation}

\noindent
This makes it clear that suppression is an important physical
phenomenon when the 2D electron gas is perturbed systematically.
Onsager's regression principle
\cite{nvk2}
immediately suggests that
screening must equally affect the stochastic perturbations
(noise) of the selfsame system. It would be very unusual 
if the thermal fluctuations were to bear no trace of $\gamma_{\rm C}$,
even as it strongly modifies the channel's differential response.

\newpage

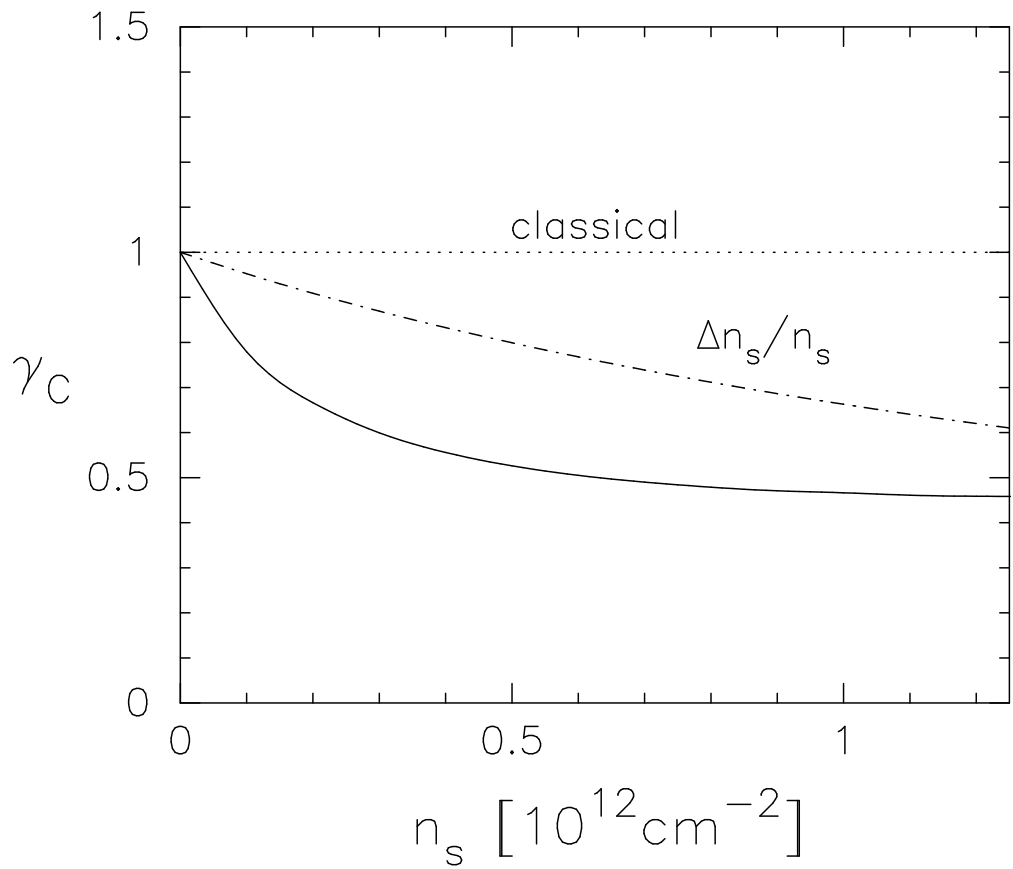
\begin{figure}
\caption{
Self-consistent Coulomb screening of a two-dimensional electron gas,
confined within a typical III-V
heterojunction quantum well at $T = 300$ K.
Solid line: the suppression factor $\gamma_{\rm C}$
for screening-induced
reduction of the electron-hole fluctuation density, below its
free value. At high density, carriers are sharply localized
in the direction transverse to current flow. Fluctuations in their
electrostatic energy provide negative feedback to reduce
the free fluctuations in density (see the Appendix).
Dash-dotted line: ratio of the free electron-hole fluctuation
density to the carrier density; the more degenerate the carriers,
the smaller the ratio. Dotted line: the classical limit for both
$\gamma_{\rm C}$ and $\Delta n_s/n_s$.
}
\label{fig1}
\end{figure}

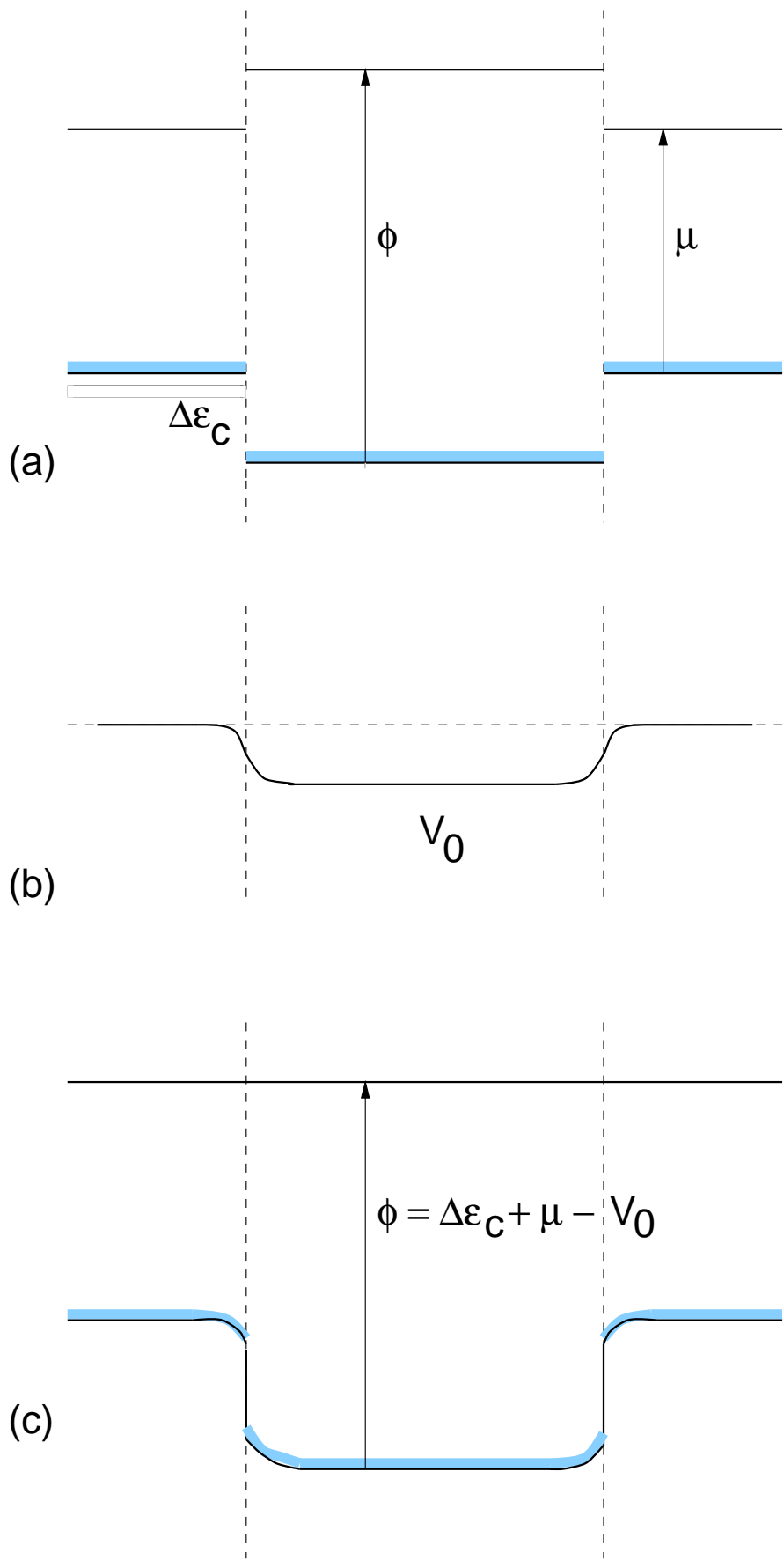
\begin{figure}
\caption{
Schematic realignment of the metallic band in a conductor terminated
by dissimilar source and drain contacts. (a) Bare conduction bands.
(b) An electrostatic contact potential $V_{\rm 0}$
develops at the interfaces. This is generated by the net transfer
of electrons, as in {\it p-n} junctions.
(c) The complete non-uniform band structure: $V_{\rm 0}$
depends on the global chemical potential $\mu$. The rigid
conduction-band offset $\Delta \varepsilon_{\rm c}$ does not.
For dense carrier populations, most of the non-uniformity
is concentrated at the interfaces.
}
\label{fig2}
\end{figure}

\begin{figure}
\caption{
The response of the local electrochemical potential $\phi$
in a non-uniform conductor-lead geometry.
(a) Complete compensation of carrier fluctuations
by the positive background. A change in global chemical
potential $\mu$ produces no change in the local electrostatic
potential $V_0$. The net change in the internal field,
due to free-carrier fluctuations, is negated
by a matching fluctuation in background density.
The free Fermi-liquid fluctuations are unchanged.
(b) Incomplete compensation.
A global change $\delta \mu$ induces a local change
$\delta V_0$ in the electrostatic potential,
partly offsetting $\delta \mu$.
Now the free Fermi-liquid fluctuations
are {\it suppressed} by interfacial Coulomb screening.
}
\label{fig3}
\end{figure}

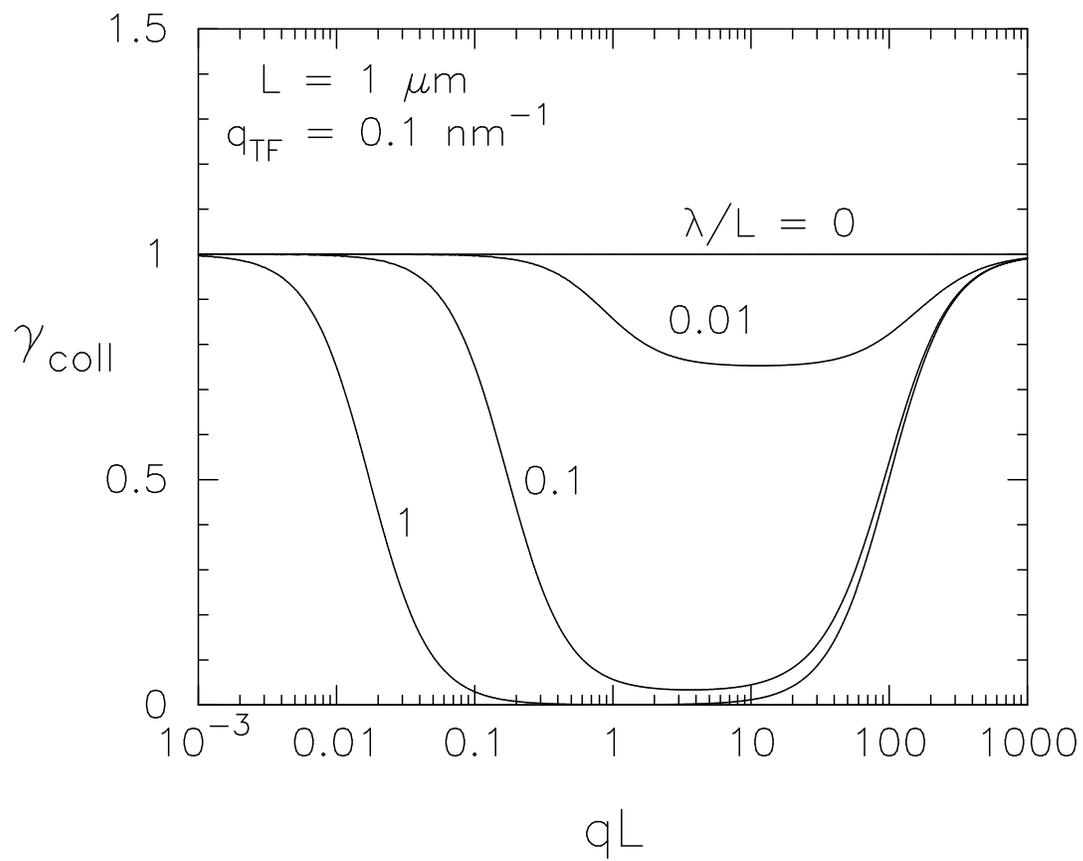
\begin{figure}
\caption{
Fourier transform of the collision-mediated screening factor
in the low-field limit of the inelastic Drude model for degenerately
doped GaAs, at density $10^{18} {\rm cm}^{-3}$.
Parameters are the Thomas-Fermi wave vector $q_{\rm TF}$,
the mean free path $\lambda$ and the sample length $L$.
For $q > q_{\rm TF}$ (short range)
the Coulomb interaction is numerically
small; there is no screening of free electron-hole fluctuations.
For $q_{\rm TF} > q > (q_{\rm TF} L \lambda)^{-1}$
(moderate range) the Coulomb interaction is stronger,
leading to appreciable collisional suppression.
For $q \ll L^{-1}$ (long range)
the Coulomb interaction is fully screened by the macroscopic leads.
Again there is no suppression. 
Collisional screening, unlike contact-potential
screening, cannot renormalize fluctuations macroscopically.
}
\label{fig4}
\end{figure}

\begin{figure}
\caption{
Schematic of a III-V heterojunction-based,
two-dimensional electron gas (2DEG)
for observing the Coulomb suppression of current noise.
(a) Plan view. The 2DEG wire is connected to large 2DEG
source and drain leads. The wire is defined by a negatively
biased, split {\it top gate}.
Uniform control of the density throughout
wire and leads is achieved by a bias voltage applied
to a {\it back gate}, beneath the channel.
The back gate should be wider than the wire to
minimize edge effects, but narrow enough to form a
resistive strip of the same length as the wire.
(b) Side view, with arrangement of back-bias and source-drain
voltages. To minimize non-uniformities in
the channel's density profile at higher currents,
the resistive back gate should support its own
driving voltage, to match the source-drain voltage
across the channel.
This offsets the local fall-off in channel potential
and keeps the effective bias constant along the wire.
}
\label{fig5}
\end{figure}


\begin{references}
\bibitem{class}
It is worth remarking that $\gamma_{\rm C} \to 1$ always in a classical
system; equipartition of energy guarantees that Maxwellian fluctuations
are not suppressed in this fashion. It is only for strongly
degenerate, interacting particles that equipartition fails
(the Fermi-Dirac distribution is not separable into purely kinetic
and purely potential factors). The screened current fluctuations are
related to fluctuations of kinetic energy, representing only
part of the total thermal energy $k_{\rm B}T$ per excitable
mode in the system; the rest is bound up in the Coulomb complement.
\bibitem{caveat}
One could not expect a wire, with uniform 2D density at equilibrium,
to remain uniform at higher fields. In principle the spatial variation
of $g$ and $\Delta f$ could be computed, with effort. More practically,
it may be easier to stabilize the local density by aligning a resistive
back gate to the 2D wire [figure 5(b)].
A drain-source potential along the gate, mirroring $V_{\rm ds}$
applied along the channel, should partly offset the local effective
bias in the channel, $V(x) \sim V_{\rm g} + xV_{\rm ds}/L$,
which would otherwise spoil the uniform density profile.
\end{references}
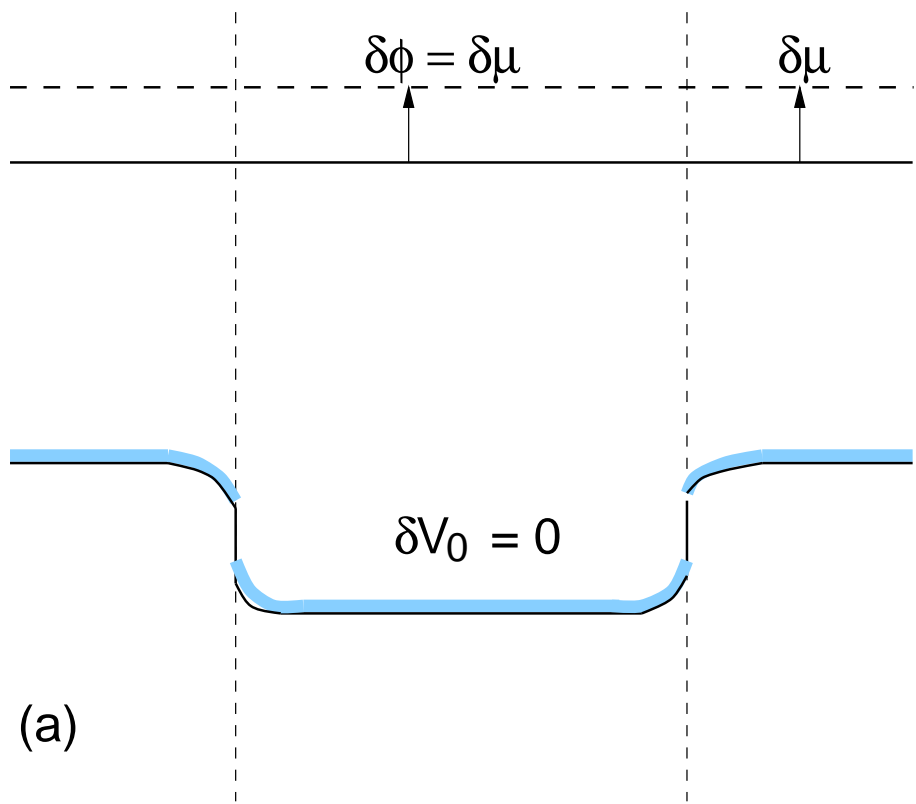
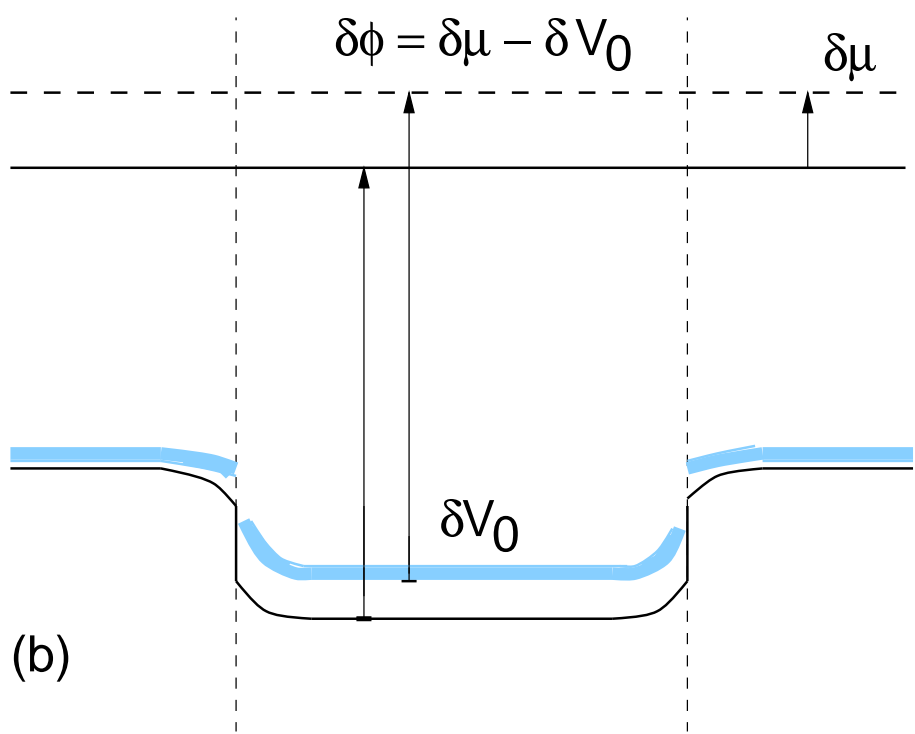
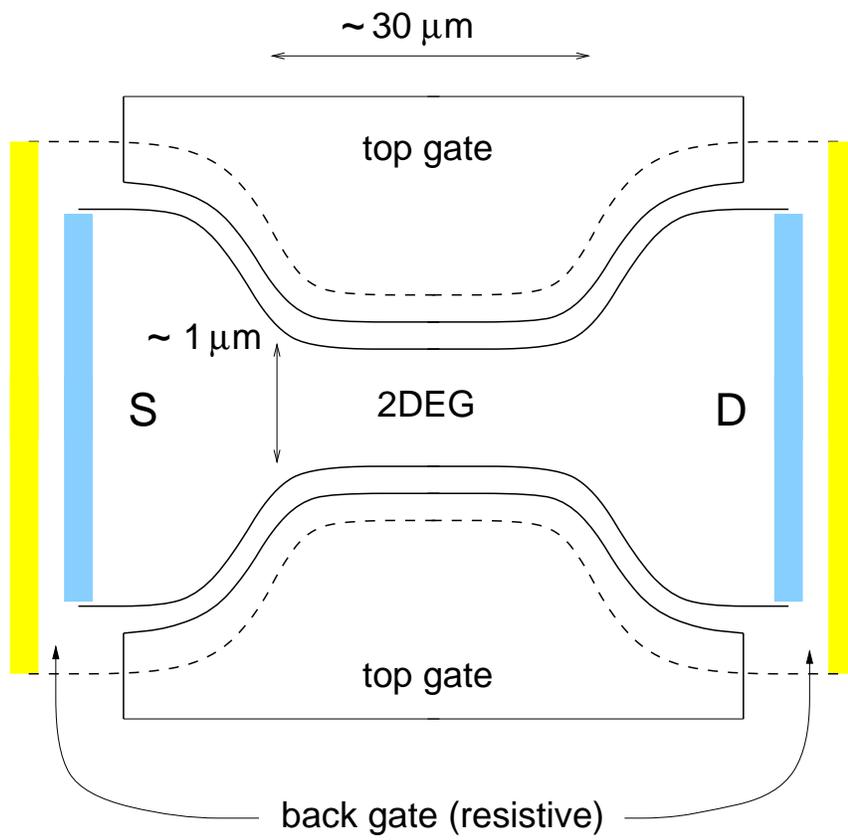
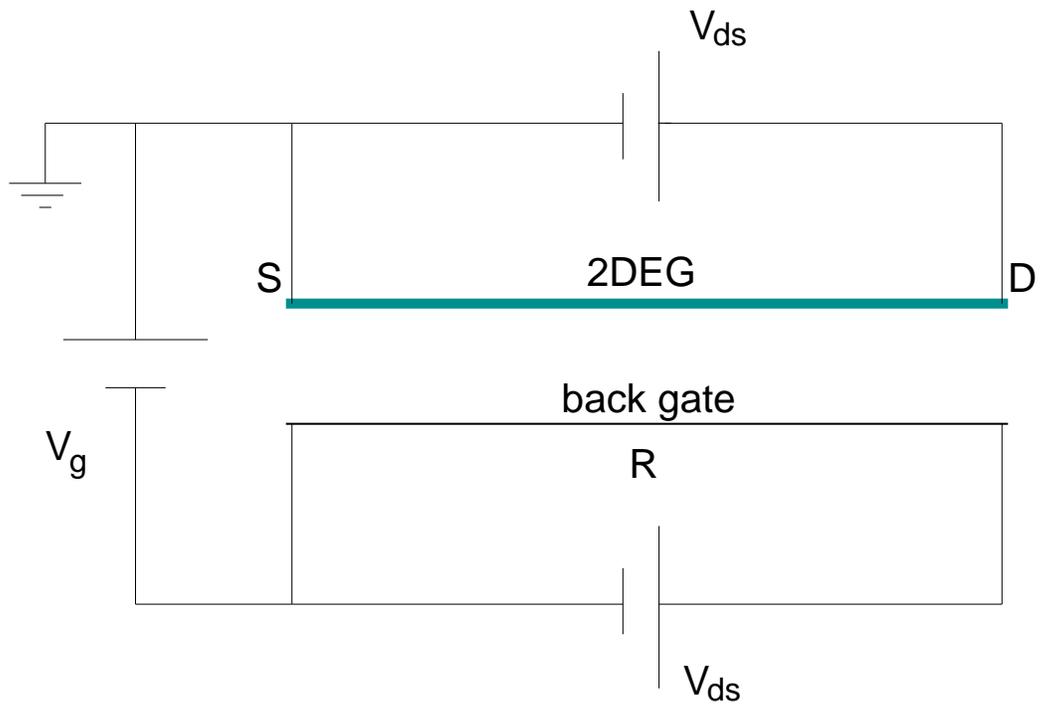
\end{document}